\documentclass[12pt,twoside]{article}

\usepackage{latexsym}
\usepackage{bbm}
\usepackage{feynmp}
\usepackage{graphicx}
\usepackage{epsfig}
\newcommand{\partialslash}{\partial\!\!\!\!\!\!\not\,\,}
\newcommand{\pslash}{p\!\!\!\!\!\not\,\,}

\newcommand{\Gslash}{G\!\!\!\!\!\!\not\,\,}
\newcommand{\Aslash}{A\!\!\!\!\!\!\not\,\,}

\begin{document}

\bibliographystyle{unsrt}
\pagestyle{empty}

\begin{flushright}
HD-THEP-00-8\\
\end{flushright}

\vspace{2cm}

Dedicated to Prof. Franz Wegner on the Occasion of his 60th Birthday

\vspace{1cm}

\begin{center}
\Large
{A Path Integral Approach to\\Inclusive Processes}\footnote{supported by the German Bundesministerium f\"ur Bildung und Forschung (BMBF), Contract no. 05 7HD 91 P(0), by Studienstiftung des Deutschen Volkes and by Deutsche Forschungsgemeinschaft (DFG), Grant no. GKR 216/1-98 }\\
\end{center}

\vspace{1cm}

\normalsize

\begin{center}
O. Nachtmann and A. Rauscher\\

\vspace{0.5cm}

Institut f\"ur Theoretische Physik\\
Universit\"at Heidelberg\\
Philosophenweg 16\\
D-69120 Heidelberg, Germany\\

\end{center}

\vspace{1cm}

\bigskip

\begin{center}
Abstract
\end{center}

\begin{quote}
The single-particle inclusive differential cross-section for a reaction $a+b\to c+X$ is written as the imaginary part of a correlation function in a forward scattering amplitude for $a+b\to a+b$ in a modified effective theory. In this modified theory the interaction Hamiltonian $\tilde H_I$ equals $H_I$ in the original theory up to a certain time. Then there is a sign change and $\tilde H_I$ becomes nonlocal. This is worked out in detail for scalar field models and for QED plus the abelian gluon model. A suitable path integral for direct calculations of inclusive cross sections is presented.
\end{quote}

\begin{fmffile}{mpbild}

\newpage

\setcounter{page}{1}

\pagestyle{plain}

\section{Introduction}

In this article we consider inclusive cross sections, i.e. reactions of the type
\begin{equation} \label{1.1}
a(p_1)+b(p_2)\longrightarrow c(p_3)+X,
\end{equation}
where $a,b,c$ are particles and $X$ stands for the unobserved remaining reaction products. We will present a general method which allows us to write the inclusive differential cross section $p^0_3d\sigma(a+b\to c+X)/d^3p_3$ as imaginary part of either a current-current or a field-field correlation function in a forward scattering amplitude $a+b\to a+b$ in a modified theory. Let
\begin{equation}\label{1.2}
H=H_0+H_I
\end{equation}
be the Hamiltonian of the original theory, with $H_0$ and $H_I$ the free and interaction parts, respectively. Then the modified theory is described by
\begin{equation}\label{1.3}
\tilde H=H_0+\tilde H_I,
\end{equation}
where $\tilde H_I$, obtained from $H_I$ in a well defined procedure, is discontinuous in time and nonlocal in space. The modified theory is constructed in such a way that both its incoming and outgoing states are equal to the incoming states of the original theory. Thus the S-matrix of the modified theory equals the unit operator. Here and in the following we always work in the Heisenberg picture of quantum mechanics.

\medskip

Our article is organised as follows. In section 2 we recall the basic relations for single inclusive cross sections. In section 3 we present our general formalism for the modified effective theory in the case of scalar fields. Quantum electrodynamics with massive photons and the theory of quarks interacting with abelian gluons are considered in section 4. We discuss some properties of the modified Hamilton operator  and derive a path integral representation for inclusive cross sections in the abelian gluon model. These techniques are then applied to the cross section $e^{+}+e^{-}\to q+X$ as a specific example. We compare our techniques with the Schwinger-Keldysh formalism \cite{1}, described e.g. in \cite{2}, and with Mueller's treatment \cite{3} of inclusive cross sections using the generalised optical theorem for $3\to 3$ scattering in section 5 which contains also our conclusions.

\section{Single Inclusive Cross Sections}

\setcounter{equation}{0}

In this section we recall some basic relations for inclusive cross sections. Our notation follows \cite{4,4a}. Let us consider a single-particle inclusive reaction, i.e
\begin{equation}\label{2.1}
a(p_{1})+b(p_{2})\to c(p_{3})+ X(p_{X}).
\end{equation}
To take a simple case, let $a,b,c$ be spinless particles with masses $m_a, m_b, m_c$. The c.m. energy squared is $s=(p_1+p_2)^2$. We assume $c(p_3)$ to differ in type and/or momentum from $a(p_1)$ and $b(p_2)$. We use the covariant normalisation for our state vectors
\begin{equation}\label{2.1a}
\langle\ a(p_1')\ |\ a(p_1)\ \rangle=(2\pi)^3\cdot 2p^0_1\cdot\delta^{(3)}(\vec{p_1}'-\vec{p}_1)
\end{equation}
and similarly for $b,c$. The $S$-matrix element for the reaction (\ref{2.1}) is given as
\begin{eqnarray}\label{2.2}
S_{fi}
&=&\langle\ c(p_{3}), X(p_{X}), out\ |\ a(p_{1}), b(p_{2}), in\ \rangle\nonumber\\
&=&iZ_{c}^{-1/2}\int d^4\!x\ e^{ixp_{3}} \langle\ X(p_{X}), out\ |\ j_{c}(x)\ |\ a(p_{1}), b(p_{2}), in\ \rangle.
\end{eqnarray}
Here we have applied the reduction formula for particle $c$ in the final state. Let $\phi_c(x)$ be a suitable interpolating field for $c$ and $Z_c$ the corresponding wave function renormalisation constant. The current $j_{c}(x)$ is defined as
\begin{equation}\label{2.3}
j_{c}(x)=\left(\Box_{x}+m^2_c\right)\phi_{c}(x).
\end{equation}
The ${\cal T}$-matrix element is obtained from the $S$-matrix element via
\begin{eqnarray}\label{2.4}
S_{fi}&=&\delta_{fi}+i(2\pi)^4\delta^{(4)}(p_{1}+p_{2}-p_{3}-p_{X}){\cal T}_{fi},\nonumber\\
{\cal T}_{fi}&=&Z_c^{-1/2}\langle\ X(p_X), out\ |\ j_c(0)\ |\ a(p_1), b(p_2), in\ \rangle.
\end{eqnarray}
The single-particle inclusive cross section $f_{inc}(p_{3})$ is defined by
\begin{eqnarray}\label{2.5}
f_{inc}(p_{3})&:=&p_{3}^0\frac{d^3\sigma}{d^3p_3}(a+b\to c+X)\nonumber\\
&\ =&\frac{1}{4(2\pi)^3w(s,m^2_a,m^2_b)}\sum_X(2\pi)^4\delta^{(4)}(p_1+p_2-p_3-p_X)|{\cal T}_{fi}|^2,
\end{eqnarray}
\begin{equation}\label{2.6}
w(x,y,z)=[x^2+y^2+z^2-2xy-2xz-2yz]^{1/2}.
\end{equation}
In the usual way the sum over all states $|X, out\ \rangle$ in (\ref{2.5}) can be carried out using completeness and translational invariance
\begin{eqnarray}\label{2.7}
f_{inc}(p_{3})
&=& \frac{1}{4(2\pi)^3w(s,m^2_a,m^2_b)}\nonumber\\
& & Z^{-1}_c\int d^4\!x\ e^{ip_3x}\langle a(p_1),b(p_2), in\ |j_c^\dagger(0)j_c(x)|a(p_1),b(p_2), in\ \rangle\\
&=& \frac{1}{2(2\pi)^3w(s,m^2_a,m^2_b)}\ \mbox{Im}\ {\cal C}(p_1,p_2,p_3),\\
{\cal C}(p_1,p_2,p_3)&=&i\int d^4\!x\ e^{ip_3x}{\cal M}(x)\label{2.8},\\
{\cal M}(x)&=&Z^{-1}_c\langle\ a(p_1), b(p_2), in\ |\ j^\dagger_c(0)j_c(x)\ |\ a(p_1), b(p_2), in\ \rangle \theta(-x^0).\qquad\label{2.9}
\end{eqnarray}
Here $\theta(z)$ is the usual step function.

\medskip

Alternatively we can write 
\begin{eqnarray}
\label{eqn6000}
{\cal M}(x)&=& Z^{-1}_c\left(\Box_{y}+m^2_c\right)\left(\Box_{z}+m^2_c\right)\nonumber\\
&&\langle a(p_1),b(p_2), in\ |\ \phi_c^\dagger(y)\phi_c(x+z)|a(p_1),b(p_2), in\ \rangle\theta(-x^0)\raisebox{-1ex}{\em $\mid_{y\to0^{-},\ z\to0^{-}}$}.\nonumber\\
\end{eqnarray}
Here the limit $y\to0^{-}$, $z\to0^{-}$ is to be understood as follows: We first require $y^0<0$ and $z^0<0$ and perform the differentiations with respect to $y$ and $z$. Afterwards we take the limit $y\to0$ and $z\to0$.

\medskip

The amplitude ${\cal C}(p_1,p_2,p_3)$ will play a central role in the following and we will be able to write it as a current-current, respectively a field-field correlation function in a forward scattering amplitude, but in a certain modified effective theory.

\section{Modified Effective Theory for Scalar Fields}

\setcounter{equation}{0}

Let us assume that the basic dynamical variables of the original theory are the operators for unrenormalised scalar fields $\phi_i(x)$ and their conjugate canonical momenta $\Pi_i(x)\ (i=1,...,N)$. For simplicity we assume that $\Pi_i(x)=\dot{\phi}_i(x)$ holds. We denote $\phi_i(x),\Pi_i(x)$ collectively as $\Phi(x)$. Let $H$ be the Hamiltonian of the system which we split into a free part $H_0$ and an interaction part $H_I$ which may depend explicitly on the time $t$, but should not involve time derivatives of $\Pi_i(x)$
\begin{equation}\label{3.1}
H(t,\Phi(\vec{x},t))=H_{0}(\Phi(\vec{x},t))+H_{I}(t,\Phi(\vec{x},t)).
\end{equation}
Besides the interacting fields and momenta $\Phi$ free fields and momenta $\Phi^{(0)}$ are considered with the corresponding Hamiltonian $H_0$. Here the mass parameters in $H_0$ are taken to be the ones of the asymptotic particles.

\medskip

We assume now as usual (cf. e.g. \cite{5,6}) that there exist\footnote{Of course this is only true in the regularised theory, i.e. for finite ultraviolet cutoff.} unitary operators $U(t)$ that realize the time-dependent canonical transformations relating $\Phi$ to $\Phi^{(0)}$
\begin{equation}\label{3.2}
\Phi(\vec{x},t)=U^{-1}(t)\Phi^{(0)}(\vec{x},t)U(t).
\end{equation}
Taking as boundary condition
\begin{equation}\label{3.2a}
\Phi(\vec x,0)=\Phi^{(0)}(\vec x,0)
\end{equation}
we get
\begin{eqnarray}\label{3.2b}
\partial_{t} U(t)&=&-iH_I(t,\Phi^{(0)}(\vec x, t))U(t),\nonumber\\
U(0)&=&{\mathbbm 1},
\end{eqnarray}
\begin{eqnarray}\label{3.3}
U(t)=\sum_{n=0}^{\infty}(-i)^{n}\int_{0}^{t}dt_{1}\int_{0}^{t_{1}}dt_{2}...\int_{0}^{t_{n-1}}dt_{n}H_{I}(t_{1})...H_{I}(t_{n}),\nonumber\\
\qquad H_{I}(t_{j})\equiv H_{I}(t_{j},\Phi^{(0)}(\vec{x},t_{j})).
\end{eqnarray}
We define furthermore
\begin{eqnarray}\label{3.5}
U(t_{2},t_{1})=U(t_{2})U^{-1}(t_{1}).
\end{eqnarray}
For $t_{2} \geq t_{1}$ we have
\begin{eqnarray}
U(t_{2},t_{1})=\mbox{T}\exp\left[-i\int_{t_{1}}^{t_{2}}dt'H_{I}(t',\Phi^{(0)}(\vec{x},t'))\right],
\end{eqnarray}
where T means time-ordering.
\medskip

Let us recall the LSZ formalism \cite{7,5,6}. Assuming for simplicity the particles $a$ and $b$ to carry the quantum numbers of some fundamental hermitian scalar fields $\phi_a, \phi_b$, we define operators
\begin{eqnarray}\label{3.6}
A(p_1,x^{0})
&=&iZ^{-1/2}_a\int d^{3}x\ e^{ip_1x}\partial_{x^{0}}\!\!\!\!\!\!\!\!\raisebox{1.5ex}{\em$\leftrightarrow$}\ \phi_a(x)\nonumber\\
&=&iZ^{-1/2}_a\int d^3x\ e^{ip_1x}(\Pi_a(x)-ip^0_1\phi_a(x)),\nonumber\\
B(p_2,x^{0})
&=&iZ^{-1/2}_b\int d^{3}x\ e^{ip_2x}\partial_{x^{0}}\!\!\!\!\!\!\!\!\raisebox{1.5ex}{\em$\leftrightarrow$}\ \phi_b(x)\nonumber\\
&=&iZ_b^{-1/2}\int d^3x\ e^{ip_2x}(\Pi_b(x)-ip^0_2\phi_b(x)),
\end{eqnarray}
where $Z_{a,b}$ are the wave function renormalisation constants. The LSZ formalism has as basic assumption that for $t\to\pm\infty$ the (hermitian conjugates of the) operators of (\ref{3.6}) converge in the weak sense to the annihilation (creation) operators of out and in-states, respectively, for instance
\begin{eqnarray}\label{3.7}
\lim_{t_{1,2}\to\pm\infty}A^\dagger(p_1,t_1) B^\dagger(p_2,t_2)\ |\ 0\ \rangle =|\ a(p_1), b(p_2)_{in}^{out}\ \rangle.
\end{eqnarray}
\medskip

Now we return to the single inclusive cross section (\ref{2.7}), where we have to calculate the matrix element (\ref{2.9})
\begin{eqnarray}\label{3.7a}
{\cal M}(x)
&=&Z_{c}^{-1}\langle\ a(p_1), b(p_2), in\ |\ j^\dagger_c(0)j_c(x)\ | \ a(p_1), b(p_2), in\ \rangle\theta(-x^0)\nonumber\\
&=&Z_{c}^{-1}\lim_{t_{i}, t_{i}'\to\infty} \langle\ 0\ |\ B(p_2,-t_{2}') A(p_1,-t_{1}')\nonumber\\
& &\qquad\qquad j^\dagger_c(0)j_c(x)A^\dagger(p_1,-t_1)B^\dagger(p_2,-t_2)\ |\ 0\ \rangle\theta(-x^0).
\end{eqnarray}
We assume the current $j_{c}(x)$ to be expressible in terms of the fields and their conjugate momenta but not involving their time derivatives. Then we have
\begin{eqnarray}\label{3.8}
U(t)j_{c}(t,\Phi(\vec{x},t))U^{-1}(t)=j_{c}(t,\Phi^{(0)}(\vec{x},t))=:j_{c}^{(0)}(x).
\end{eqnarray}
Furthermore we have from (\ref{3.2}) and (\ref{3.6})
\begin{eqnarray}\label{3.9}
U(t)A(p_1,t;\Phi(\vec x,t))U^{-1}(t)=A(p_1,t; \Phi^{(0)}(\vec x,t))=:A^{(0)}(t)\end{eqnarray}
and the same for $B(p_2,t)$. We note that the $\Phi^{(0)}$ satisfy the free field equations. In the case of a free scalar field with mass $m$ they read
\begin{eqnarray}\label{3.10}
\dot\phi^{(0)}(\vec x,t)&=&\Pi^{(0)}(\vec x,t),\nonumber\\
\dot\Pi^{(0)}(\vec x,t)&=&(\Delta_x-m^2)\phi^{(0)}(\vec x,t).
\end{eqnarray}
This implies
\begin{eqnarray}\label{3.11}
\dot A^{(0)}(t)=0,\quad\dot B^{(0)}(t)=0.
\end{eqnarray}
Now ${\cal M}(x)$ (\ref{3.7a}) may be written as
\begin{eqnarray}\label{3.12}
{\cal M}(x)&=&
\lim_{t_{i},t_{i}'\to\infty}Z_{c}^{-1}\langle\ 0\ |\ U^{-1}(-T')\nonumber\\
& & \qquad\qquad U(-T',-t_2')B^{(0)}(-t_2')U(-t_2',-t_1')A^{(0)}(-t_1')\nonumber\\
& & \qquad\qquad U(-t_1',0)j_c^{\dagger(0)}(0)U(0,x^0)j_c^{(0)}(x)U(x^0,-t_1)\nonumber\\
& & \qquad\qquad A^{\dagger(0)}(-t_1)U(-t_1,-t_2)B^{\dagger(0)}(-t_2)U(-t_2,-T)\nonumber\\
& & \qquad\qquad\qquad\qquad\qquad\qquad U(-T)\ |\ 0\ \rangle\theta(-x^0).
\end{eqnarray}
Here we have introduced further times $T, T'$ and we assume without loss of generality
\begin{eqnarray}\label{3.13}
& & T> t_2>t_1,\nonumber\\& & T'>t_2'>t_1'.
\end{eqnarray}
With the usual assumption that the interaction is switched off adiabatically for $t\to\pm\infty$ we get from the adiabatic theorem
\begin{eqnarray}\label{3.14}
\lim_{T\to\infty}U(-T)\ |\ 0\ \rangle=e^{i\varphi_-} |\ 0\ ),
\end{eqnarray}
where $|0)$ is the vacuum state of the free theory. Inserting everything in (\ref{3.12}), we get
\begin{eqnarray}\label{3.15}
{\cal M}(x)&=&\lim_{t_{i},t_{i}'\to\infty}\lim_{T,T'\to\infty}Z_{c}^{-1}\nonumber\\
& & \qquad\qquad(\ 0\ |\ U(-T',-t_2')B^{(0)}(-t_2')U(-t_2',-t_1')A^{(0)}(-t_1')\nonumber\\
& & \qquad\qquad U(-t_1',0)j_c^{\dagger(0)}(0)U(0,x^0)j_c^{(0)}(x)U(x^0,-t_1)\nonumber\\
& & \qquad\qquad A^{\dagger(0)}(-t_1)U(-t_1,-t_2)B^{\dagger(0)}(-t_2)U(-t_2,-T)|\ 0\ )\theta(-x^0)\nonumber\\
& & \qquad\qquad [(\ 0\ |\ U(-T')U^{-1}(-T)\ |\ 0\ )]^{-1}.
\end{eqnarray}
Following the time-dependence in ${\cal M}(x)$ from the right to the left, we start at time $-T\to-\infty$ and pass through operators of increasing time arguments until time $0$. Then the time sequence changes and we go back in time to time $-T'\to-\infty$. In a usual matrix element the time arguments should increase instead.

\medskip

We will now show that we can write the matrix element ${\cal M}(x)$ (\ref{3.15}) in the usual form, with time increasing from right to left, if we pass to operators $\tilde{U}$ of the form (\ref{3.3}) but with a modified interaction Hamiltonian $\tilde{H}_{I}$. As the time-sequence in ${\cal M}(x)$ is correct up to $t=0$, we request
\begin{eqnarray}\label{3.16}
\tilde{U}(t)=\theta(-t)U(+t)+\theta(t)U(-t).
\end{eqnarray}
This gives with (\ref{3.2b})
\begin{eqnarray}\label{3.17}
\partial_{t}\tilde U(t)=-i\tilde H_I(t)\tilde U(t),
\end{eqnarray}
\begin{eqnarray}\label{3.18}
\tilde{H}_{I}(t)&=&\theta(-t)H_{I}(t,\Phi^{(0)}(\vec{x},t))-\theta(t)H_{I}(-t,\Phi^{(0)}(\vec{x},-t)).
\end{eqnarray}
For $t>0$ our modified interaction Hamiltonian $\tilde H_I(t)$ depends on the free fields and momenta at time $(-t)$. But we know how to express the free fields and momenta at time $(-t)$ by their values at time $t$ using the free field equations of motion. For $\phi^{(0)}, \Pi^{(0)}$ satisfying (\ref{3.10}) we get
\begin{eqnarray}\label{3.19}
\phi^{(0)}(\vec x,x^0)&=&\int_{y^0=const.}d^3y\ \Delta(x-y;m^2)\partial_{y^{0}}\!\!\!\!\!\!\!\!\raisebox{1.5ex}{\em$\leftrightarrow$}\ \phi^{(0)}(y)\nonumber\\
&=&\int_{y^0=const.}d^3y\left\{\Delta(x-y;m^2)\Pi^{(0)}(y)+\dot{\Delta}(x-y;m^2)\phi^{(0)}(y)\right\},\nonumber\\[2mm]
\Pi^{(0)}(\vec x,x^0)&=&\dot{\phi}^{(0)}(\vec x,x^0)
\end{eqnarray}
with the usual commutator function for scalar fields of mass $m$
\begin{eqnarray}\label{3.20}
\Delta(z;m^2)&=&i\int d^{R}k (e^{-ikz}-e^{ikz}),\nonumber\\
d^{R}k &=&\frac{d^4k}{(2\pi)^3}\theta(k^0)\delta(k^2-m^2).
\end{eqnarray}
Setting $y^0=t$ and $x^0=-t$ in (\ref{3.19}) we get expressions for $\phi^{(0)}(\vec x,-t)$ and $\Pi^{(0)}(\vec x,-t)$ as linear functions of $\phi^{(0)}(\vec y,t)$ and $\Pi^{(0)}(\vec y,t)$. Thus, we can consider $\tilde H_I(t)$ (\ref{3.18}) as a \underbar{nonlocal} functional of the dynamical variables $\Phi^{(0)}(\vec x,t)$ at the \underbar{same} time $t$ also for $t>0$
\begin{eqnarray}\label{3.21}
\tilde H_I(t)=\tilde H_I(t,\Phi^{(0)}(\vec x,t)).
\end{eqnarray}
Using in addition (\ref{3.11}) we can rewrite ${\cal M}(x)$ (\ref{3.15}) as follows
\begin{eqnarray}\label{3.22}
{\cal M}(x)
&=&\lim_{t_{i},t_{i}'\to\infty}\lim_{T,T'\to\infty}Z_{c}^{-1}\nonumber\\
& & (\ 0\ |\ \tilde U(T',t_2')B^{(0)}(t_2')\tilde U(t_2',t_1')A^{(0)}(t_1')\nonumber\\
& & \qquad\tilde U(t_1',0)j_c^{\dagger(0)}(0)\tilde U(0,x^0)j_c^{(0)}(x)\tilde U(x^0,-t_1)\nonumber\\
& & \qquad A^{\dagger(0)}(-t_1)\tilde U(-t_1,-t_2)B^{\dagger(0)}(-t_2)\tilde U(-t_2,-T)\ |\ 0\ )\theta(-x^0)\nonumber\\
& & \qquad\qquad\qquad [(\ 0\ |\ \tilde U(T',-T)\ |\ 0\ )]^{-1},
\end{eqnarray}
\begin{eqnarray}\label{3.23}
\tilde U(t',t)&=&\tilde U(t')\tilde U^{-1}(t)=\mbox{T}\exp\left[-i\int^{t'}_t
dt''\tilde H_I(t'',\phi^{(0)}(\vec x,t''))\right]\quad (t'\geq t).\qquad
\end{eqnarray}
In the limit $T,T'\to\infty$ we get from (\ref{3.22})
\begin{eqnarray}\label{3.24}
{\cal M}(x)
&=&\lim_{t_{i}t_{i}'\to\infty}Z_{c}^{-1}(\ 0\ |\ \mbox{T} \exp\left[-i\int^\infty_{-\infty}dt'\tilde H_I(t')\right]\nonumber\\
& &\qquad B^{(0)}(t_2')A^{(0)}(t_1')j^{\dagger(0)}_c(0)j_c^{(0)}(x) A^{\dagger(0)}(-t_1)B^{\dagger(0)}(-t_2)\ |\ 0\ )\theta(-x^0)\nonumber\\
& &\qquad\qquad\qquad \left[(\ 0\ |\ \mbox{T} \exp\left[-i\int^\infty_{-\infty}dt'\tilde H_I(t')\right]\ |\ 0\ )\right]^{-1}.
\end{eqnarray}
Clearly, we can consider (\ref{3.24}) as matrix element of the standard type but in the modified theory governed by the total Hamiltonian
\begin{eqnarray}\label{3.25}
\tilde H(t,\tilde\Phi(\vec x,t))=H_0(\tilde\Phi(\vec x,t))+\tilde
H_I(t,\tilde\Phi(\vec x,t)),
\end{eqnarray}
\begin{eqnarray}\label{3.26}
\tilde\Phi(\vec x,t)&=&\tilde U^{-1}(t)\Phi^{(0)}(\vec x,t)\tilde U(t),
\end{eqnarray}
\begin{eqnarray}\label{3.27}
{\cal M}(x)=Z_{c}^{-1}\langle\!\langle\ a(p_1), b(p_2), out\ |\ j_c^\dagger(0) j_c(x)\ |\ a(p_1), b(p_2), in\ \rangle\!\rangle \theta(-x^0).
\end{eqnarray}
Here we denote by $\langle\!\langle\ \rangle\!\rangle$ matrix elements in the modified theory. Using (\ref{eqn6000}) we can also write ${\cal M}$ as 
\begin{eqnarray}\label{eqn7000}
{\cal M}(x)&=&Z_{c}^{-1}\left(\Box_{y}+m^2_c\right)\left(\Box_{z}+m^2_c\right)\nonumber\\
&&\langle\!\langle\ a(p_1), b(p_2), out\ |\ \phi_c^\dagger(y) \phi_c(x+z)\ |\ a(p_1), b(p_2), in\ \rangle\!\rangle \theta(-x^0)\raisebox{-1ex}{\em $\mid_{y\to0^{-},\ z\to0^{-}}$}.\nonumber\\
\end{eqnarray}

\medskip

The S-matrix of the modified theory equals unity, due to the fact that the incoming and outgoing states are now identical
\begin{eqnarray}
\label{eqn8000}
\tilde{S}={\mathbbm 1}.
\end{eqnarray}

\medskip

In (\ref{3.25}) - (\ref{eqn7000}) we have a main result of our paper. The matrix element ${\cal M}$ is written in the standard form with an in-state to the right and an out-state to the left. The prize we have to pay is that we have to use the modified theory where the Hamiltonian $\tilde{H}(t)$ has a sudden variation at $t=0$ and is nonlocal for $t>0$. On the other hand we can use (\ref{3.25}) - (\ref{eqn7000}) as starting point to write down a path integral representation for ${\cal M}$ in the standard way. Below in section 4 we will do this for QED and the abelian gluon model as an example.

\section{Inclusive Production in QED and an Abelian Gluon Model}

\setcounter{equation}{0}

\subsection{QED and the Abelian Gluon Model}

In this section inclusive reactions are considered in QED coupled to an abelian gluon model. An example for such a reaction is
\begin{eqnarray}
\label{eqn90}
e^{+}(p_{1})+e^{-}(p_{2})\ \to\ q(p_{3})+X.
\end{eqnarray}

\medskip

Starting point is the Lagrangian describing the interaction of electrons of mass $m$ and charge $-e$ with a massive photon of mass $\lambda$ - to avoid any infrared divergences - and of two quark flavours of equal mass $M$ and electric charge $eQ_{q}$ with the photon and a massive abelian gluon of mass $\mu$. As Lagrangian we choose
\begin{eqnarray}
\label{eqn1000}
{\cal L}&=&-\ \frac{1}{4}G_{\mu\nu}G^{\mu\nu}-\frac{1}{2\eta_{0}}(\partial_{\mu}G^{\mu})^2+\frac{1}{2}\mu_{0}^2G_{\mu}G^{\mu}\nonumber\\
&&-\ \frac{1}{4}F_{\mu\nu}F^{\mu\nu}-\frac{1}{2\xi_{0}}(\partial_{\mu}A^{\mu})^2+\frac{1}{2}\lambda_{0}^2A_{\mu}A^{\mu}\nonumber\\
&&+\ \bar{\psi}\left(\frac{i}{2}\!\!\!\!\stackrel{\ \ \leftrightarrow}{\partialslash}-m_{0}+e_{0}\Aslash\ \right)\psi+\bar{q}\left(\frac{i}{2}\!\!\!\!\stackrel{\ \ \leftrightarrow}{\partialslash}-M_{0}-e_{0}Q_{q}\Aslash-g_{0}\tau_{3}\Gslash\ \right)q,
\end{eqnarray}
where $G_{\mu}$ denotes the abelian gluon field and $G_{\mu\nu}=\partial_{\mu}G_{\nu}-\partial_{\nu}G_{\mu}$ its field strength tensor, $A_{\mu}$ the photon field and $F_{\mu\nu}=\partial_{\mu}A_{\nu}-\partial_{\nu}A_{\mu}$ its field strength tensor, $\psi$ the electron field and $q$ the quark field. For the vector boson part of ${\cal L}$ we have chosen here a form advocated by Stueckelberg (cf. e.g.\cite{6}), where terms $-1/2\eta_{0}(\partial_{\mu}G^{\mu})^2$ and $-1/2\xi_{0}(\partial_{\mu}A^{\mu})^2$ have been added for the abelian gluon and the photon field. We have a quark field with two flavours and $\tau_{3}$ is the usual Pauli matrix. All quantities in (\ref{eqn1000}) are the unrenormalised ones. The corresponding renormalised quantities are denoted either without any subscript in the case of constants or with the subscript $R$ in the case of fields. 

\medskip

The Lagrangian (\ref{eqn1000}) has two discrete symmetries of the charge conjugation type, $C_{A}$ and $C_{G}$, which guarantee that no further coupling terms need to be added in (\ref{eqn1000}) and that photons and the abelian gluons do not mix. All this is explained in appendix A, where we also discuss the renormalisation procedure in our model. 

\medskip

The canonical momenta of the vector fields of the theory are listed in (\ref{eqn9000}) of appendix B. For finite $\xi_{0}$ and $\eta_{0}$ the relation (\ref{eqn9000}) can be solved for $\partial_{0}A_{\mu}$ and $\partial_{0}G_{\mu}$ and the canonical formalism can be applied.

\medskip

The Hamiltonian corresponding to (\ref{eqn1000}) can be obtained as usual by a Legendre transformation
\begin{eqnarray}
H&=&H_{0}(y^0)+H_{I}(y^0),\nonumber\\
H_{0}(y^0)&=&\int d^3y{\cal H}_{0}(y),\nonumber\\
H_{I}(y^0)&=&\int d^3y{\cal H}_{I}(y),
\end{eqnarray}
where we choose the free ${\cal H}_{0}(y)$ and the interaction part ${\cal H}_{I}(y)$ of the Hamiltonian density as follows
\begin{eqnarray}
\label{eqn24000}
{\cal H}_{0}(y)
&=&-\frac{\eta}{2}\Pi_{G0}\Pi_{G}^{0}-\frac{1}{2}\Pi_{Gi}\Pi_{G}^{i}-\partial_{i}G^{i}\Pi_{G}^0+\Pi_{G}^{i}\partial_{i}G^0+\frac{1}{4}G_{ij}G^{ij}-\frac{1}{2}\mu^2G_{\mu}G^{\mu}\qquad\nonumber\\
&&-\frac{\xi}{2}\Pi_{A0}\Pi_{A}^{0}-\frac{1}{2}\Pi_{Ai}\Pi_{A}^{i}-\partial_{i}A^{i}\Pi_{A}^0+\Pi^{i}\partial_{i}A^0+\frac{1}{4}A_{ij}A^{ij}-\frac{1}{2}\lambda^2A_{\mu}A^{\mu}\nonumber\\
&&-\bar{\psi}\left(\frac{i}{2}\gamma^{i}\!\!\!\!\stackrel{\ \ \leftrightarrow}{\ \partial_{i}}-m\right)\psi-\bar{q}\left(\frac{i}{2}\gamma^{i}\!\!\!\!\stackrel{\ \ \leftrightarrow}{\ \partial_{i}}-M\right)q.\\
{\cal H}_{I}(y)\label{23000}
&=&+\frac{1}{2}\delta\mu^2G_{\mu}G^{\mu}+\frac{1}{2}\delta\eta\Pi_{G0}\Pi_{G}^{0}+\frac{1}{2}\delta\lambda^2A_{\mu}A^{\mu}+\frac{1}{2}\delta\xi\Pi_{A0}\Pi_{A}^{0}\nonumber\\
&&-\delta m\bar{\psi}\psi-\delta M\bar{q}q\nonumber\\
&&-e_{0}\bar{\psi}\Aslash\psi+\bar{q}\left(e_{0}Q_{q}\Aslash+g_{0}\tau_{3}\Gslash\ \right)q,
\end{eqnarray}
with
\begin{eqnarray}
\delta\mu^2&=&\mu^2-\mu_{0}^2,\nonumber\\
\delta\eta&=&\eta-\eta_{0},\nonumber\\
\delta\lambda^2&=&\lambda^2-\lambda_{0}^2,\nonumber\\
\delta\xi&=&\xi-\xi_{0},\nonumber\\
\delta m&=&m-m_{0},\nonumber\\
\delta M&=&M-M_{0}.
\end{eqnarray}
Only the interaction term has to be altered in order to construct the Hamilton operator for the modified theory according to the rules specified in section 3 
\begin{eqnarray}\label{eqn16}
\tilde{H}(y^0)&=&H_{0}(y^0)+\tilde{H}_{I}(y^0),\nonumber\\
\tilde{H}_{I}(y^0)&=&\int d^3y\ [\ \theta(-y^0){\cal H}_{I}(y)-\theta(y^0)\tilde{\cal H}_{I}(\tilde{y})\ ]\nonumber\\
&&\qquad\qquad\mbox{with}\ \tilde{y}=(-y^0,\vec{y}).
\end{eqnarray}
In $\tilde{\cal H}_{I}(\tilde{y})$ the fields at time $-y^0$ have to be substituted by the fields at time $y^0$ via the time shift relations analogous to (\ref{3.19}). For the vector fields and the fermion fields these are given by (\ref{eqn50}), (\ref{eqn80}) and (\ref{eqn60}) in appendix B. Two points should be stressed concerning this modified Hamilton operator: First, due to the jump in the interaction, the effective theory has no time translation invariance, so that no longer energy conservation holds at every vertex. Second, the effective interaction contains derivatives. Thus, if we use perturbation theory for the evaluation of amplitudes in the effective theory, we have to be careful about contact terms in time ordered products, especially for the vector fields (for a general discussion of such contact terms see \cite{6,6a}).

\subsection{Inclusive Quark Production in $e^{+}e^{-}$-Annihilation}

We want to calculate the single-quark-inclusive cross section of reaction (\ref{eqn90}). Performing all the steps as described for scalar fields in section 3 we can write this inclusive cross section as imaginary part of an amplitude ${\cal C}$ ( cf. (\ref{2.7}, \ref{2.8}, \ref{eqn6000})). For unpolarised $e^{-}$ and $e^{+}$ in the initial state and summation over spin and flavour of the final state quark this reads 

\begin{eqnarray}
f_{inc}(p_{3})&=&\frac{1}{2(2\pi)^3w(s,m^2,m^2)}\ \mbox{Im}\ {\cal C}(p_1,p_2,p_3),\label{eqn12000}\\
{\cal C}(p_1,p_2,p_3)&=&i\int d^4\!x\ e^{ip_3x}{\cal M}(x),\\
{\cal M}(x)&=&-Z^{-1}_{q}\sum_{s_{q}}\bar{u}(p_{3})(i\overrightarrow{\partialslash_{z}}-M)\sum_{s_{e^{+}},s_{e^{-}}}\ \!\!\!\!\!\!'\nonumber\\
&&\ \langle e^{+}(p_{1}),e^{-}(p_{2}),in |\ \mbox{T}q_{A}(x+z)\bar{q}_{A}(y)\ |\ e^{+}(p_{1}),e^{-}(p_{2}),in \rangle\theta(-x^0)\nonumber\\
&&\quad\qquad\qquad\qquad\qquad\qquad\qquad\qquad(-i\overleftarrow{\partialslash_{y}}-M)u(p_{3})\raisebox{-1ex}{\em $\mid_{y\to 0^{-},\ z\to 0^{-}}$}.\nonumber\\ \label{eqna}\
\end{eqnarray}
The transition to the effective theory gives
\begin{eqnarray}\label{eqn11000}
{\cal M}(x)&=&-Z^{-1}_{q}\sum_{s_{q}}\bar{u}(p_{3})(i\overrightarrow{\partialslash_{z}}-M)\nonumber\\
&&\sum_{s_{e^{+}},s_{e^{-}}}\ \!\!\!\!\!\!'\!\langle\!\langle\ e^{+}(p_{1}),e^{-}(p_{2})\ |\ \mbox{T}q_{A}(x+z)\bar{q}_{A}(y)\ |\ e^{+}(p_{1}),e^{-}(p_{2}) \rangle\!\rangle\theta(-x^0)\nonumber\\
&&\quad\qquad\qquad\qquad\qquad\qquad\qquad\qquad(-i\overleftarrow{\partialslash_{y}}-M)u(p_{3})\raisebox{-1ex}{\em $\mid_{y\to 0^{-},\ z\to 0^{-}}$}.\nonumber\\ \label{eqnb}
\end{eqnarray}
Here $\sum\ \!\!\!\!'$ means the average over the spin states, $A$ is the quark flavour index over which we sum. In (\ref{eqna}) and (\ref{eqnb}) we have for convenience inserted a T-product symbol which puts $q$ and $\bar{q}$ into the order required for $x^0<0$ in the limit $y\to0^{-}$.

\medskip

In order to calculate ${\cal M}(x)$ we expand it in powers of the electromagnetic coupling constant $e$. The leading term in the connected part of  ${\cal M}$ is of order $e^4$, corresponding to the diagrams of figure 1 which can be classified as $e^{+}e^{-}$-annihilation (a) and $e^{+}e^{-}$-scattering (b) ones.

\begin{figure}[ht]
\begin{center}\unitlength1mm
\begin{eqnarray}
\parbox{60mm}{(a)\begin{fmfgraph*}(60,30)
\fmfsurroundn{v}{4}
\fmffreeze
\fmfleft{i1,i2}\fmfright{o1,o2}
\fmf{electron,label=$e^{-}$,label.side=left}{v6,i1}\fmf{electron,label=$e^{+}$}{i2,v6}
\fmf{electron,label=$e^{-}$,label.side=left}{o1,v5}\fmf{electron,label=$e^{+}$}{v5,o2}
\fmf{photon}{v5,v7}\fmf{photon}{v7,v6}\fmfv{decor.shape=circle,decor.filled=hatched,decor.size=25}{v7}\fmf{quark,label=$q$}{v4,v7}\fmf{quark,label=$q$,label.side=left}{v7,v2}\fmfv{decor.shape=circle,decor.filled=shaded,decor.size=12}{v5,v6}
\end{fmfgraph*}}\qquad\qquad
\parbox{40mm}{(b)\begin{fmfgraph*}(40,50)
\fmfsurroundn{v}{4}
\fmffreeze
\fmfleft{i1,i2}\fmfright{o1,o2}
\fmf{electron,label=$e^{-}$}{v6,i1}\fmf{electron,label=$e^{+}$}{i2,v5}
\fmf{electron,l.side=right,label=$e^{-}$}{o1,v6}\fmf{electron,label=$e^{+}$}{v5,o2}
\fmf{photon}{v5,v7}\fmf{photon}{v7,v6}\fmfv{decor.shape=circle,decor.filled=hatched,decor.size=25}{v7}\fmf{quark,label=$q$}{v7,v3}\fmf{quark,label=$q$}{v1,v7}\fmfv{decor.shape=circle,decor.filled=shaded,decor.size=12}{v5,v6}
\end{fmfgraph*}}
\nonumber
\end{eqnarray}
\label{fig1}
\caption{Annihilation diagrams (a) and scattering diagrams (b) for the amplitude ${\cal M}(x)$ in  order $e^4$. In the middle blob arbitrary gluonic interactions are allowed. All lines and vertices are in the effective theory. Time runs from right to left.}
\end{center}
\end{figure}
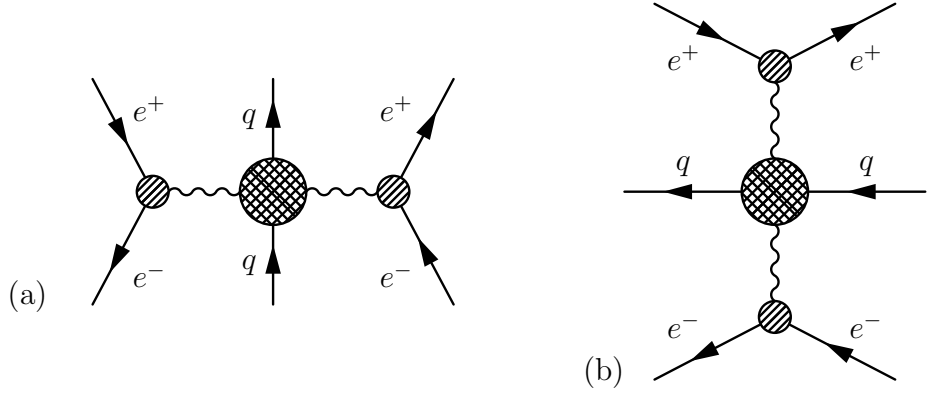

\medskip

Using the LSZ reduction formalism we get for the connected part of the annihilation matrix element ${\cal M}^{a}$ 
\begin{eqnarray}
&&\sum_{s_{e^{+}},s_{e^{-}}}\ \!\!\!\!\!\!'\langle\!\langle e^{+}(p_{1}),e^{-}(p_{2}) |\ \mbox{T}q_{A}(x+z)\bar{q}_{A}(y)\ |\ e^{+}(p_{1}),e^{-}(p_{2}) \rangle\!\rangle=\nonumber\\
&&\qquad=-e^2 l^{\mu\nu}\langle\!\langle 0 |\ \mbox{T}\int\! d^4\!x_{3}[\theta(-x_{3}^0)A_{\mu}(x_{3})e^{i(p_{1}+p_{2})x_{3}}-\theta(x_{3}^0)A_{\mu}(\tilde{x_{3}})e^{i(p_{1}+p_{2})\tilde{x_{3}}}]\nonumber\\
&&\qquad\qquad\qquad\qquad\qquad\qquad q_{A}(x+z)\bar{q}_{A}(y)\nonumber\\
&&\qquad\qquad\qquad\int\! d^4\!x_{4}[\theta(-x_{4}^0)A_{\nu}(x_{4})e^{-i(p_{1}+p_{2})x_{4}}-\theta(x_{4}^0)A_{\nu}(\tilde{x_{4}})e^{-i(p_{1}+p_{2})\tilde{x_{4}}}]\ | 0 \rangle\!\rangle,\nonumber\\
\end{eqnarray}
where we have written the matrix element with the help of the lepton tensor
\begin{eqnarray}
l^{\mu\nu}&=&\sum_{s_{e^{+}},s_{e^{-}}}\ \!\!\!\!\!\!'\bar{u}(p_{2})\gamma^{\mu}v(p_{1})\bar{v}(p_{1})\gamma^{\nu}u(p_{2})\nonumber\\
&=&p_{1}^{\mu}p_{2}^{\nu}+p_{1}^{\nu}p_{2}^{\mu}-g^{\mu\nu}(p_{1}p_{2}+m^2).
\end{eqnarray}
Expanding further we get to lowest order in $Q_{q}e_{0}$
\begin{eqnarray}\label{eqn500000}
&&\sum_{s_{e^{+}},s_{e^{-}}}\ \!\!\!\!\!\!'\langle\!\langle e^{+}(p_{1}),e^{-}(p_{2}) |\ \mbox{T}q_{A}(x+z)\bar{q}_{A}(y)\ |\ e^{+}(p_{1}),e^{-}(p_{2}) \rangle\!\rangle\nonumber\\
&&\qquad=\frac{1}{2}e^4Q_{q}^2l^{\mu\nu}\nonumber\\
&&\qquad\qquad\langle\!\langle 0 |\ \mbox{T}\int\! d^4\!x_{3}[\theta(-x_{3}^0)A_{\mu}(x_{3})e^{i(p_{1}+p_{2})x_{3}}-\theta(x_{3}^0)A_{\mu}(\tilde{x_{3}})e^{i(p_{1}+p_{2})\tilde{x_{3}}}]\nonumber\\
&&\qquad\qquad\qquad\int\! d^4\!x_{1}[\theta(-x_{1}^0)\bar{q}(x_{1})\Aslash(x_{1})q(x_{1})-\theta(x_{1}^0)\bar{q}(\tilde{x}_{1})\Aslash(\tilde{x}_{1})q(\tilde{x}_{1})]\nonumber\\
&&\qquad\qquad\qquad\qquad\qquad\qquad q_{A}(x+z)\bar{q}_{A}(y)\nonumber\\
&&\qquad\qquad\qquad\int\! d^4\!x_{2}[\theta(-x_{2}^0)\bar{q}(x_{2})\Aslash(x_{2})q(x_{2})-\theta(x_{2}^0)\bar{q}(\tilde{x}_{2})\Aslash(\tilde{x}_{2})q(\tilde{x}_{2})]\nonumber\\
&&\qquad\qquad\qquad\int\! d^4\!x_{4}[\theta(-x_{4}^0)A_{\nu}(x_{4})e^{-i(p_{1}+p_{2})x_{4}}-\theta(x_{4}^0)A_{\nu}(\tilde{x_{4}})e^{-i(p_{1}+p_{2})\tilde{x_{4}}}]\ | 0 \rangle\!\rangle=\nonumber\\
&&\qquad=-e^4Q_{q}^2l^{\mu\nu}\frac{1}{s-\lambda^2-i\epsilon}\frac{1}{s-\lambda^2+i\epsilon}\nonumber\\
&&\langle\!\langle 0 |\ \mbox{T}\int\!\! d^4\!x_{1}[\theta(-x_{1}^0)e^{i(p_{1}+p_{2})x_{1}}\bar{q}(x_{1})\gamma_{\mu}q(x_{1})-\theta(x_{1}^0)e^{i(p_{1}+p_{2})\tilde{x}_{1}}\bar{q}(\tilde{x}_{1})\gamma_{\mu}q(\tilde{x}_{1})]\nonumber\\
&&\qquad\qquad\qquad\qquad\qquad\qquad q_{A}(x+z)\bar{q}_{A}(y)\nonumber\\
&&\qquad\int\!\! d^4\!x_{2}[\theta(-x_{2}^0)e^{-i(p_{1}+p_{2})x_{2}}\bar{q}(x_{2})\gamma_{\nu}q(x_{2})-\theta(x_{2}^0)e^{-i(p_{1}+p_{2})\tilde{x}_{2}}\bar{q}(\tilde{x}_{2})\gamma_{\nu}q(\tilde{x}_{2})] | 0 \rangle\!\rangle,\nonumber\\
\end{eqnarray}
so that our matrix element ${\cal M}^{a}(x)$ can be written in the following way
\begin{eqnarray}
\label{eqn20000}
&&{\cal M}^{a}(x)=Z^{-1}_{q}e^4Q_{q}^2l^{\mu\nu}\left\vert\frac{1}{s-\lambda^2+i\epsilon}\right\vert^2\nonumber\\
&&\ \sum_{s_{q}}\bar{u}(p_{3})(i\overrightarrow{\partialslash_{z}}-M)\theta(-x^0)\langle\!\langle 0 |\mbox{T}q_{A}(x+z)\nonumber\\
&&\qquad\int\!\! d^4\!x_{2}[\theta(-x_{2}^0)e^{-i(p_{1}+p_{2})x_{2}}\bar{q}(x_{2})\gamma_{\nu}q(x_{2})-\theta(x_{2}^0)e^{-i(p_{1}+p_{2})\tilde{x}_{2}}\bar{q}(\tilde{x}_{2})\gamma_{\nu}q(\tilde{x}_{2})]\nonumber\\
&&\qquad\qquad\int\!\! d^4\!x_{1}[\theta(-x_{1}^0)e^{i(p_{1}+p_{2})x_{1}}\bar{q}(x_{1})\gamma_{\mu}q(x_{1})-\theta(x_{1}^0)e^{i(p_{1}+p_{2})\tilde{x}_{1}}\bar{q}(\tilde{x}_{1})\gamma_{\mu}q(\tilde{x}_{1})]\nonumber\\
&&\qquad\qquad\qquad\qquad\qquad\qquad\qquad\qquad\bar{q}_{A}(y) | 0 \rangle\!\rangle(-i\overleftarrow{\partialslash_{y}}-M)u(p_{3})\raisebox{-1ex}{\em $\mid_{y\to 0^{-},\ z\to 0^{-}}$}\nonumber\\
\end{eqnarray}
in terms of quark 6-point-functions in the quark-gluon-theory. In the next subsection this will be represented by a path integral in the modified abelian gluon model. The electromagnetic interaction of the incoming fermions could be separated, the calculation is sketched in Appendix C.

\medskip

The same procedure can be applied to the scattering matrix element ${\cal M}^{b}(x)$ corresponding to the diagram of figure 1b. Of course this diagram cannot give a contribution to the inclusive cross section. In higher orders the $e^{+}e^{-}$-scattering type diagrams give e.g. the contributions of 2-photon annihilation processes (cf. figure 2) to the inclusive cross section (\ref{eqn90})
\begin{eqnarray}
e^{+}+e^{-}\ \to\ e^{+}+e^{-}+q+X.
\end{eqnarray}

\medskip

\begin{figure}[ht]
\begin{center}\unitlength1mm
\begin{eqnarray}
\parbox{40mm}{(a)\begin{fmfgraph*}(40,50)
\fmfsurroundn{v}{24}
\fmffreeze
\fmfleft{i1,i2}\fmfright{o1,o2}
\fmf{electron,label=$e^{-}$}{v25,i1}\fmf{electron,l.side=right,label=$e^{+}$}{i2,v26}
\fmf{electron,l.side=right,label=$e^{-}$}{o1,v25}\fmf{electron,label=$e^{+}$}{v26,o2}\fmf{photon}{v25,v27}\fmf{photon}{v27,v26}\fmfv{decor.shape=circle,decor.filled=empty,decor.size=25}{v27}\fmfdot{v25,v26}
\fmffreeze
\fmf{quark,label=$q$}{v27,v15}\fmf{plain}{v27,v12}\fmf{plain}{v27,v11}\fmf{plain}{v27,v13}\fmfv{label=$X$}{v12}
\end{fmfgraph*}}
\qquad\qquad
\parbox{50mm}{(b)\begin{fmfgraph*}(60,40)
\fmfsurroundn{v}{4}
\fmfleft{i1,i2}\fmfright{o1,o2}
\fmf{electron,label=$e^{-}$}{v5,i1}\fmf{electron}{v6,v5}\fmf{electron,label=$e^{-}$}{o1,v6}
\fmf{electron,label=$e^{+}$}{i2,v7}\fmf{electron}{v7,v8}\fmf{electron,label=$e^{+}$}{v8,o2}
\fmfv{decor.shape=circle,decor.filled=shaded,decor.size=12}{v7,v8,v6,v5}
\fmffreeze\fmf{phantom}{v9,v11}\fmf{phantom}{v11,v10}\fmf{phantom}{v9,v11}\fmf{phantom}{v11,v10}\fmf{phantom}{v9,v11}\fmf{phantom}{v11,v10}
\fmf{photon}{v5,v10}\fmf{photon}{v6,v9}\fmf{photon}{v7,v10}\fmf{photon}{v8,v9}
\fmfv{decor.shape=circle,decor.filled=hatched,decor.size=60}{v11}\fmf{quark,label=$q$}{v1,v9}\fmf{quark,label=$q$}{v10,v3}
\end{fmfgraph*}}
\nonumber
\end{eqnarray}
\label{fig2}
\caption{The diagrams of 2-photon annihilation processes for inclusive quark production (\ref{eqn90}) are shown in (a), the corresponding diagrams for ${\cal M}$ (\ref{eqn11000}) in (b). Time runs from right to left.}
\end{center}
\end{figure}
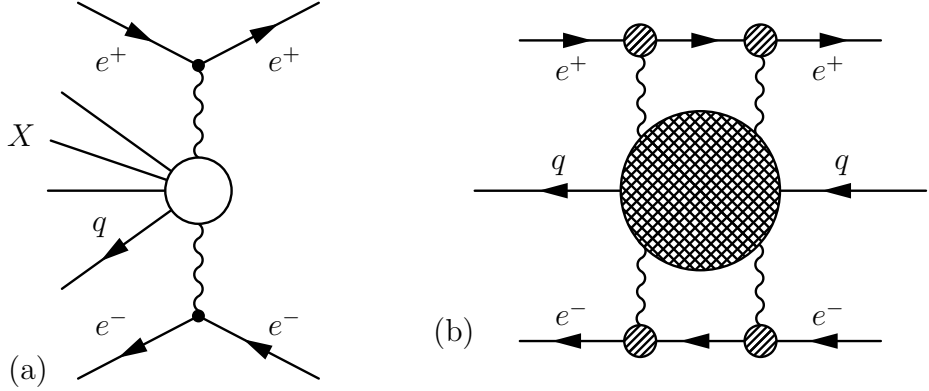

As a simple check of our theoretical manipulations let us finally calculate the matrix element ${\cal M}^{a}(x)$ in lowest order in the quark-gluon-coupling $g$, i.e. for $g=0$. Decomposing the quark 6-point-functions with Wick's theorem into quark 2-point-functions and substituting for these the perturbative propagators (\ref{eqn100}) we get for ${\cal M}^{a}$ two contributions according to the two different possible contractions
\begin{eqnarray}
{\cal M}^{a}(x)&=&2e^4Q_{q}^2l^{\mu\nu}\left\vert\frac{1}{s-\lambda^2+i\epsilon}\right\vert^2\theta(-x^0)\nonumber\\
&&\qquad\int\!d^{R}p \sum_{s_{q}}\bar{u}(p_{3})\ [\ e^{i(p-p_{1}-p_{2})x}\gamma_{\nu}(\pslash-M)\gamma_{\mu}\nonumber\\
&&\qquad\qquad\qquad\qquad\qquad +\ e^{i(p+p_{1}+p_{2})x}\gamma_{\mu}(\pslash-M)\gamma_{\nu}\ ]\ u(p_{3}).
\end{eqnarray}
Inserting this in (\ref{eqn12000}) we get

\begin{eqnarray}
f_{inc}(p_{3})&=&\frac{1}{2(2\pi)^3w(s,m^2,m^2)}\pi e^4Q_{q}^2 l^{\mu\nu}\left\vert\frac{1}{s-\lambda^2+i\epsilon}\right\vert^2\nonumber\\
& &2\sum_{s_{q}}\bar{u}(p_{3})\gamma_{\nu}(\pslash-M)\gamma_{\mu}u(p_{3})\delta^{(1)}(p^2-M^2)\theta(p^0)\raisebox{-1ex}{\em $\mid_{p=p_{1}+p_{2}-p_{3}}$}.
\end{eqnarray}
This is, of course, the standard result, which one obtains in considering to lowest order the reaction

\begin{eqnarray}
e^{+}+e^{-}\to q+\bar{q}
\end{eqnarray}  
for two quark flavours of charge $Q_{q}$.

\subsection{Path Integral Representation}

In the last subsection the matrix element ${\cal M}(x)$ and therefore the one-quark-inclusive cross-section could be expressed in terms of quark 6-point-functions after separating the electromagnetic interaction by a perturbative calculation. As the coupling $g$ is not assumed to be small we will now derive a representation of these quark 6-point-functions suitable for non-perturbative calculations. For this we consider the Hamiltonian path integral in our effective theory obtained from the abelian gluon model.

\medskip

Any Green's function of the theory can be written as

\begin{eqnarray}
&&\langle\!\langle 0 |\ \mbox{T}q(x_{1})... \bar{q}(x_{2})... G^{\lambda}(x_{3})... \Pi_{G}^{\rho}(x_{4})...\ | 0 \rangle\!\rangle\nonumber\\
&&\qquad\qquad ={\tilde{\cal Z}}^{-1}\int{\cal D}(G,\Pi_{G},q,\bar{q})\ q(x_{1})...\bar{q}(x_{2})...G^{\lambda}(x_{3})...\Pi_{G}^{\rho}(x_{4})...\nonumber\\
&&\qquad\qquad\qquad\qquad\qquad\times\exp \{i\int\! d^4\!y\ (\Pi(y)\dot{\Phi}(y)-\tilde{{\cal H}}(y))\}
\end{eqnarray}
with

\begin{eqnarray}
\label{eqn22000}
{\tilde{\cal Z}}&=&\int{\cal D}(G,\Pi_{G},q,\bar{q})\ \exp \{i\int\! d^4\!y\ (\Pi(y)\dot{\Phi}(y)-\tilde{{\cal H}}(y))\} 
\end{eqnarray}
and 

\begin{eqnarray}
\Pi(y)\dot{\Phi}(y)&=&\Pi_{G}^{\rho}(y)\dot{G}_{\rho}(y)+\Pi_{q}(y)\dot{q}(y)+\dot{\bar{q}}(y)\Pi_{\bar{q}}(y)\nonumber\\
&=&\Pi_{G}^{\rho}(y)\dot{G}_{\rho}(y)+\frac{i}{2}\bar{q}(y)\gamma_{0}\!\!\!\!\stackrel{\ \ \leftrightarrow}{\ \ \partial_{0}}q(y). 
\end{eqnarray}
Here the classical fields $G$ and $\Pi_{G}$ and the Grassmann fields $q$ and $\bar{q}$ have to be inserted into the part of $\tilde{\cal H}(y)$ (\ref{eqn16}) which describes the abelian gluon model. Furthermore for $y^0>0$ the classical and Grassmann fields have to be time-shifted according to (\ref{eqn50}) and (\ref{eqn60}) so that we get

\begin{eqnarray}
&&\Pi(y)\dot{\Phi}(y)-\tilde{{\cal H}}(y)=\nonumber\\
&&\qquad\frac{1}{2}\eta_{0}\Pi_{G0}(y)\Pi_{G}^0(y)+\partial_{\mu}G^{\mu}(y)\Pi_{G}^0(y)+\frac{1}{2}\Pi_{Gi}(y)\Pi_{G}^{i}(y)-G_{i0}(y)\Pi_{G}^{i}(y)\nonumber\\
& &\qquad\qquad-\frac{1}{4}G_{ij}(y)G^{ij}(y)+\frac{1}{2}\lambda_{0}^2G_{\mu}(y)G^{\mu}(y)+\bar{q}(y)\left(\frac{i}{2}\!\!\!\!\!\stackrel{\ \ \leftrightarrow}{\partialslash}_{y}-M_{0}\right)q(y)\nonumber\\
& &\qquad\qquad\qquad-\ \theta(-y^0)\ g_{0}\ \bar{q}(y)\ \tau_{3}\Gslash(y)\ q(y)\nonumber\\
& &\qquad\qquad\qquad\qquad\qquad+\ \theta(y^0)\ g_{0}\ \bar{q}(\tilde{y})\ \tau_{3}\Gslash(\tilde{y})\ q(\tilde{y}). 
\end{eqnarray}

\medskip

For the modified abelian gluon model the canonical momenta $\Pi_{G}^{\mu}$ can be integrated out explicitely to obtain a Lagrangian path integral 

\begin{eqnarray}
&&\int{\cal D}(G,\Pi_{G},q,\bar{q})\ \exp \{i\int\! d^4\!y\ (\Pi(y)\dot{\Phi}(y)-\tilde{{\cal H}}(y))\}\nonumber\\
&&\qquad\qquad=\int{\cal D}(G,q,\bar{q})\ \exp \{i\int\! d^4\!y\ \tilde{\cal L}(y)\} 
\end{eqnarray}
with an effective Lagrangian $\tilde{\cal L}(y)$

\begin{eqnarray}
\tilde{\cal L}(y)&=&
-\ \frac{1}{4}G_{\mu\nu}(y)G^{\mu\nu}(y)-\frac{1}{2\eta_{0}}(\partial_{\mu}G^{\mu}(y))^2+\frac{1}{2}\mu_{0}^2G_{\mu}(y)G^{\mu}(y)\nonumber\\
&&\qquad+\ \bar{q}(y)(\frac{i}{2}\!\!\!\!\!\stackrel{\ \ \leftrightarrow}{\partialslash}_{\!y}-M_{0})q(y)\nonumber\\
& &\qquad\qquad-\theta(-y^0)g_{0}\bar{q}(y)\tau_{3}\Gslash(y)q(y)+\theta(y^0)g_{0}\bar{q}(\tilde{y})\tau_{3}\Gslash(\tilde{y})q(\tilde{y})\nonumber\\
&&-\theta(y^0)\frac{1}{2}g^2_{0}\int\!d^4\!z_{1}\ \delta^{(1)}(y^{0}-z_{1}^0)\int\!d^4\!z_{2}\ \delta^{(1)}(y^{0}-z_{2}^0)\bar{q}(\tilde{z_{1}})\tau_{3}\nonumber\\
&&\ \ \{\frac{1}{\eta_{0}}(-\Delta(\tilde{z_{1}}-y;\lambda_{0}^2)\gamma_{0}+\frac{1}{\lambda_{0}^2}\partialslash\ \partial_{0}[\Delta(\tilde{z_{1}}-y;\eta_{0}\lambda_{0}^2)-\Delta(\tilde{z_{1}}-y;\lambda_{0}^2)])q(\tilde{z_{1}})\nonumber\\
&&\qquad\bar{q}(\tilde{z_{2}})(-\Delta(\tilde{z_{1}}-y;\lambda_{0}^2)\gamma^0+\frac{1}{\lambda_{0}^2}\partialslash\ \partial^0[\Delta(\tilde{z_{1}}-y;\eta_{0}\lambda_{0}^2)-\Delta(\tilde{z_{1}}-y;\lambda_{0}^2)])\nonumber\\
&&\ \ +(-\Delta(\tilde{z_{1}}-y;\lambda_{0}^2)\gamma_{i}+\frac{1}{\lambda_{0}^2}\partialslash\ \partial_{i}[\Delta(\tilde{z_{1}}-y;\eta_{0}\lambda_{0}^2)-\Delta(\tilde{z_{1}}-y;\lambda_{0}^2)])q(\tilde{z_{1}})\nonumber\\
&&\qquad\bar{q}(\tilde{z_{2}})(-\Delta(\tilde{z_{1}}-y;\lambda_{0}^2)\gamma^{i}+\frac{1}{\lambda_{0}^2}\partialslash\ \partial^{i}[\Delta(\tilde{z_{1}}-y;\eta_{0}\lambda_{0}^2)-\Delta(\tilde{z_{1}}-y;\lambda_{0}^2)])\}\nonumber\\
&&\qquad\qquad\qquad\qquad\qquad\qquad\qquad\qquad\qquad\qquad\qquad\qquad\tau_{3}q(\tilde{z_{2}}).
\end{eqnarray}
The straightforward calculation leads to a four fermion coupling term, which results from the time-shift relations of the bosonic fields.

\medskip
   
As the Hamiltonian path integral (\ref{eqn22000}) is quadratic in $q,\bar{q}$, the fermionic fields can be integrated out. With the Green's function $\tilde{S}_{F}$ of the quark in the modified gluon background
\begin{eqnarray}
(i\partialslash-M_{0})\tilde{S}_{F}(z_{1},z_{2};G)-g\int\! d^4\!z\ \Gslash(z_{1};z)\tilde{S}_{F}(z,z_{2};G)=-\delta^{(4)}(z_{1}-z_{2}),
\end{eqnarray}
the quark propagator is given as
\begin{eqnarray}
\label{eqn25000}
\langle\!\langle\ 0\mid Tq(z_{1})\bar{q}(z_{2})\mid 0\rangle\!\rangle=\langle\!\langle\ \frac{1}{i}\tilde{S}_{F}(z_{1},z_{2};G)\rangle\!\rangle,
\end{eqnarray}
where the brackets $\langle\!\langle\ \rangle\!\rangle$ on the right hand side denote the average over all gluon fields with the measure dictated by the path-integral
\begin{eqnarray}
\langle\!\langle F(G,\Pi_{G})\rangle\!\rangle&=&{\tilde{\cal Z}}^{-1}\int{\cal D}(G,\Pi_{G})F(G,\Pi_{G})\nonumber\\
& &\qquad\times\exp[\ i\int\! d^4\!y\ (\frac{1}{2}\eta_{0}\Pi_{G0}\Pi_{G}^0+\partial_{\mu}G^{\mu}\Pi_{G}^0+\frac{1}{2}\Pi_{Gi}\Pi_{G}^{i}-G_{i0}\Pi_{G}^{i}\nonumber\\
& &\qquad\qquad\qquad\qquad\qquad\qquad-\frac{1}{4}G_{ij}G^{ij}+\frac{1}{2}\lambda_{0}^2G_{\mu}G^{\mu})]\nonumber\\
& &\qquad\qquad\times\left(\mbox{det}[-i\left(i\partialslash-M_{0}-g_{0}\int\! d^4\!z \Gslash(\ .;z)\right)]\right)^2
\end{eqnarray} 
with
\begin{eqnarray}
\Gslash(y_{1};y_{2})&=&\theta(-y_{1}^0)\delta^{(4)}(y_{1}-y_{2})\ \tau_{3}\Gslash(y_{1})\nonumber\\
& &+\theta(y_{1}^0)\delta^{(1)}(y_{1}^0-y_{2}^0)\int\! d^4\!z\ \delta(z^0-y_{1}^0)\nonumber\\
&&\qquad\qquad\qquad\qquad\qquad \gamma^0 S(y_{1}-\tilde{z};M)\tau_{3}\Gslash(\tilde{z})S(\tilde{z}-y_{2};M)\gamma^0.\nonumber\\
\end{eqnarray}
Inserting the Green's function (\ref{eqn25000}) into the matrix element (\ref{eqn20000}) we get the matrix elements and therefore the inclusive cross section expressed explicitely in terms of a Hamiltonian or Lagrangian path integral, respectively. These expressions should be a convenient starting point for applying non-perturbative methods to an evaluation of inclusive cross sections. It should be possible, for instance, to generalise the methods of \cite{9} to the case of this effective theory here.

\section{Comparison to Other Techniques and Conclusions}

\setcounter{equation}{0}

In this section we give a brief comparison of our technique using time-shifted fields with the formalisms of Schwinger and Keldysh \cite{1,2} and Mueller \cite{3}. Then we will draw our conclusions.

\medskip

To discuss the Schwinger-Keldysh formalism we go back to (\ref{3.12}) where the time sequence runs forward from $-T$ to $0$ and then back to $-T'$. Following Schwinger and Keldysh one inserts $U$-matrices from 0 to a large positive time $T''$ and back to $0$

\begin{eqnarray}
U(0,T'')U(T'',0)={\mathbbm 1}, 
\end{eqnarray}

\begin{eqnarray}
{\cal M}(x)
&=&\lim_{t_{i},t_{i}'\to\infty}\lim_{T,T',T''\to\infty}Z_{c}^{-1}\nonumber\\
& & (\ 0\ |\ U(-T',-t_2')B^{(0)}(-t_2') U(-t_2',-t_1')A^{(0)}(-t_1')\nonumber\\
& & \qquad U(-t_1',0)U(0,T'')U(T'',0)j_c^{\dagger(0)}(0)U(0,x^0)j_c^{(0)}(x)U(x^0,-t_1)\nonumber\\
& & \qquad A^{\dagger(0)}(-t_1)U(-t_1,-t_2)B^{\dagger(0)}(-t_2)U(-t_2,-T)\ |\ 0\ )\theta(-x^0)\nonumber\\
& & \qquad\qquad\qquad [(\ 0\ |\ U(T',-T)\ |\ 0\ )]^{-1}.
\end{eqnarray}
In the complex time plane we go along a path starting at $-T$, going to  $T''$ and then back to $-T'$ (figure 3). This method is best used introducing the two field formalism \cite{1,2}. It plays an important role in thermal field theory in the real time formulation \cite{11,12}.

\begin{figure}[ht]
\begin{center}
  \hfil\\
  \begin{fmfgraph*}(100,40)
    \fmfiset{arrow_ang}{10}
    \fmfipair{minf,mthreem,mthreep,zerom,zerop,threem,threep,inf}
    \fmfiset{minf}{(0,.5h)}
    \fmfiset{inf}{(w,.5h)}
    \fmfi{plain}{minf--inf}
    \fmfi{plain}{(.5w,0)--(.5w,h)}
    \fmfiv{l=$\mathop{\mathop{\textrm{t-plane}}}$,l.angle=90}{(.95w,1h)}
    \fmfpen{thick}
    \fmfiset{zerom}{(.5w,.4h)}
    \fmfiset{zerop}{(.5w,.6h)}
    \fmfiset{threem}{(.9w,.4h)}
    \fmfiset{threep}{(.9w,.6h)}
    \fmfiset{mthreem}{(.1w,.4h)}
    \fmfiset{mthreep}{(.1w,.6h)}
    \fmfi{electron}{mthreep--zerop}
    \fmfi{plain}{zerop--threep}
    \fmfi{plain}{threem--zerom}
    \fmfi{electron}{zerom--mthreem}
    \fmfiv{l=$T''$,l.angle=45}{(threep)}
    \fmfiv{l=$-T$,l.angle=135}{(mthreep)}
    \fmfiv{l=$-T'$,l.angle=-135}{(mthreem)}
    \fmfi{plain,right=0.5}{threep--threem}
    \end{fmfgraph*}
\end{center}
\label{fig3}
\caption{Time contour in the Keldysh formalism.}
\end{figure}
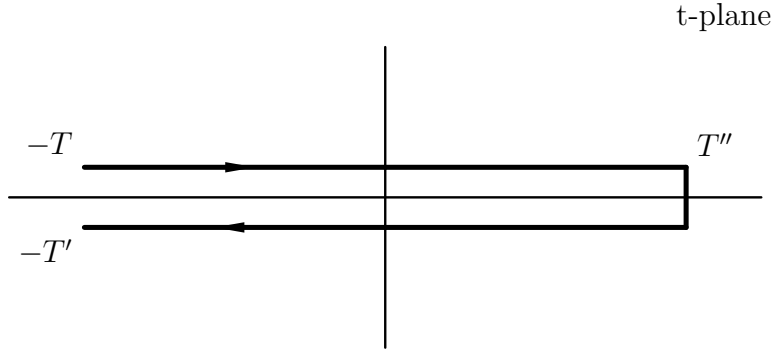

\medskip

In Mueller's formalism one starts with the 3-3 amplitude for $a+b+\bar{c}\to a+b+\bar{c}$
\begin{eqnarray}
&&\langle\ a(p_{1}'),b(p_{2}'),\bar{c}(p_{3}'), out\ |\ a(p_{1}),b(p_{2}),\bar{c}(p_{3}), in\ \rangle=\nonumber\\
&&\qquad\qquad=\delta_{fi}+i(2\pi)^4\delta^{(4)}(p_{1}+p_{2}+p_{3}-p_{1}'-p_{2}'-p_{3}'){\cal T}_{fi}.
\end{eqnarray}
Then one identifies a particular contribution to Im${\cal T}_{fi}$, i.e. a particular discontinuity of ${\cal T}_{fi}$ (figure 4).
\begin{figure}[ht]
\begin{center}\unitlength1mm
\begin{eqnarray}
\parbox{40mm}{\begin{fmfgraph*}(30,20)
  \fmfleft{i1,i2,i3} \fmfright{o1,o2,o3}
  \fmf{fermion}{v1,i1}\fmf{fermion}{v1,i2}
  \fmf{fermion}{v1,o3}\fmf{fermion}{i3,v1}
  \fmf{fermion}{o1,v1}\fmf{fermion}{o2,v1}
  \fmfv{decor.shape=circle,decor.filled=empty,decor.size=20}{v1}
  \fmflabel{a}{i1}\fmflabel{a}{o1}\fmflabel{b}{i2}
  \fmflabel{b}{o2}\fmfv{label=$\bar{c}$}{i3}
  \fmfv{label=$\bar{c}$}{o3}
  \fmfi{dashes}{(.5w,0)--(.5w,h)}
\end{fmfgraph*}}\qquad\qquad
\parbox{60mm}{\begin{fmfgraph*}(60,20)
  \fmfleft{i1,i2}\fmfright{o1,o2}
  \fmftop{v1,v2,v3,v4,v7,v8}
  \fmf{fermion,label=a}{v5,i1}\fmf{fermion,label=b}{v5,i2}
  \fmf{fermion,label=a}{o1,v6}\fmf{fermion,label=b}{o2,v6}
  \fmf{fermion,label=c}{v3,v5}\fmf{fermion,label=c}{v6,v4}
  \fmf{phantom}{i1,v5}\fmf{phantom}{v6,o1}
  \fmf{phantom}{i1,v5}\fmf{phantom}{v6,o1}
  \fmf{vanilla,left=-0.2,tension=0.5}{v5,v6}
  \fmf{vanilla,left=-0.5,tension=0.5}{v5,v6}
  \fmf{vanilla,left=-0.35,tension=0.5}{v5,v6}
  \fmf{vanilla,left=-0.05,tension=0.5}{v5,v6}
  \fmf{vanilla,left=0.15,tension=0.5,l.s=left,label=$\ \ \ \ X$}{v5,v6}  
  \fmfv{decor.shape=circle,decor.filled=empty,decor.size=20}{v5} 
  \fmfv{decor.shape=circle,decor.filled=empty,decor.size=20}{v6}
  \fmfi{dashes}{(.5w,0)--(.5w,h)}
\end{fmfgraph*}}\nonumber
\end{eqnarray}
\end{center}
\label{fig4}
\caption{In (a) the discontinuity of ${\cal T}_{fi}$ is shown which is related to the inclusive cross section $a+b\to c+X$ in Mueller's formalism (b).}
\end{figure}
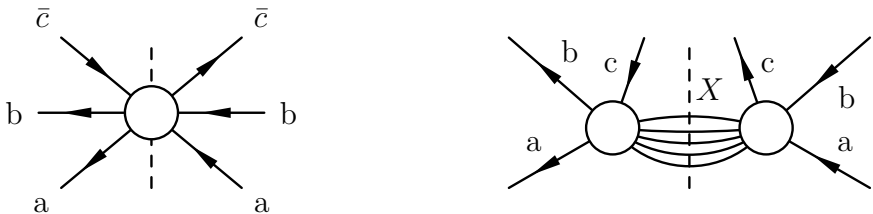

\medskip

We should emphasise that our approach is quite distinct from these. In our effective theory the time contour runs from $-\infty$ to $+\infty$. The scattering matrix $\tilde{S}={\mathbbm 1}$ (cf. \ref{eqn8000}). If we insert the matrix element ${\cal M}$ calculated in the effective theory according to (\ref{eqn7000}) into (\ref{2.8}) we get an expression which looks almost like a 3$\to$3 scattering amplitude but not quite so because of the function $\theta(-x^0)$ in (\ref{eqn7000}).

\medskip

To conclude: We have presented in this article a novel technique for treating inclusive reactions: the method of the time-shifted fields. We have explained this method for theories with scalar fields and with electrons and quarks interacting with photons and abelian gluons. The generalisation to QCD should be straightforward and will be dealt with in future work. We have written the inclusive cross section as imaginary part of an amplitude ${\cal C}$ for which we have given a path integral representation. We have applied our formalism to the reaction $e^{+}+e^{-}\to q+X$ and checked that in lowest nontrivial order we get the correct result. Our hope is that our path integral representation will lead to a genuinely non-perturbative evaluation of inclusive cross sections at high energies along the lines of \cite{9}. The method of the time shifted fields could also be useful for general studies of inclusive reactions in QFT. One can, for instance, think of generalising Wilson's operator product expansion (OPE) \cite{13} to our effective theory. Then our methods should allow a straightforward description of inclusive production of hadrons $h$ in $e^{+}e^{-}$ annihilation at high energies
\begin{eqnarray}
e^{+}+e^{-}\to h+X
\end{eqnarray}
in terms of a genuine OPE. A similar approach should be possible for fracture functions \cite{14} for hadron-hadron or virtual photon ($\gamma^*$)-proton scattering
\begin{eqnarray}
h_{1}+h_{2}&\to&h_{3}+X\nonumber\\
\gamma^*+p&\to&p+X.
\end{eqnarray}
The last reaction is, of course, related to diffractive deep inelastic scattering as observed at HERA \cite{15}. Finally, our methods can also be applied to the treatment of non-equilibrium processes in thermal field theory. 

\section*{Acknowledgements}

The authors would like to thank E. Berger, W. Bernreuther and P.V. Landshoff for discussions and H.G. Dosch and A. Hebecker for reading the manuscript and for useful suggestions and F. Cooper for correspondance.
 
\appendix

\section*{Appendix}

\section{Relations for QED and the Abelian Gluon\\ Model}

\setcounter{equation}{0}

\subsection{Symmetries}

The Lagrangian (\ref{eqn1000}) has the following two symmetries of the charge conjugation type
\begin{eqnarray}
C_{A}:\qquad \psi(x) &\to& {\cal C}\bar{\psi}^{T}(x),\label{eqn36000}\qquad\nonumber\\
q(x) &\to &\epsilon{\cal C}\bar{q}^{T}(x),\nonumber\\
A^{\mu}(x) &\to& -A^{\mu}(x),\nonumber\\
G^{\mu}(x) &\to& G^{\mu}(x);\\
C_{G}:\qquad \psi(x)&\to& \psi(x),,\label{eqn37000}\qquad\nonumber\\
q(x) &\to& -\epsilon q(x),\nonumber\\
A^{\mu}(x) &\to& A^{\mu}(x),\nonumber\\
G^{\mu}(x) &\to& -G^{\mu}(x).
\end{eqnarray}
Here ${\cal C}$ is the usual charge conjugation matrix for Dirac fields (cf. e.g. \cite{10}) and
\begin{eqnarray}
\epsilon=\left(\begin{array}{rr}0&1\\-1&0\end{array}\right).
\end{eqnarray} 
The standard charge conjugation transformation is
\begin{eqnarray}
C=C_{A}C_{G}.
\end{eqnarray}
The symmetries $C_{A}$ and $C_{G}$ forbid coupling terms of the types $AG$, $A^3$, $AG^2$, $A^2G$, $G^3$, $AG^3$, $A^3G$ in our Lagrangian (\ref{eqn1000}) and require the quark mass term to be flavour diagonal. Coupling terms of type $A^4$, $A^2G^2$, $G^4$ have mass dimension 4 and could be included in ${\cal L}$. But we can leave them out since the renormalisation procedure does not require them. The reason is as in QED: The vector bosons couple to conserved currents and thus the superficial degree of divergence of vertex functions of four vector bosons is less than 0 (cf. e.g. \cite{5}).

\subsection{Renormalisation}

We outline here the main steps in the renormalisation program of our theory.

\medskip

Let us define the unrenormalised vacuum polarisation tensors $\Pi^{(A)}_{\mu\nu}(k)$ of the photon and $\Pi^{(G)}_{\mu\nu}(k)$ of the gluon
\begin{figure}[ht]
\begin{center}\unitlength1mm
\begin{eqnarray}
\parbox{20mm}{\begin{fmfgraph*}(20,10)\fmfleft{i1}\fmfright{o1}
  \fmf{photon}{v1,i1}\fmfv{decor.shape=circle,decor.filled=hatched,decor.size=20}{v1}\fmf{photon,label=$\leftarrow k$}{o1,v1}\fmflabel{$\mu$}{i1}\fmflabel{$\nu$}{o1}   
\end{fmfgraph*}}\qquad&=&ie_{0}^2\Pi^{(A)}_{\mu\nu}(k),\\
\nonumber\\
\parbox{20mm}{\begin{fmfgraph*}(20,10)\fmfleft{i1}\fmfright{o1}
  \fmf{gluon}{v1,i1}\fmfv{decor.shape=circle,decor.filled=hatched,decor.size=20}{v1}\fmf{gluon,label=$\leftarrow k$}{o1,v1}\fmflabel{$\mu$}{i1}\fmflabel{$\nu$}{o1}   
\end{fmfgraph*}}\qquad&=&ig_{0}^2\Pi^{(G)}_{\mu\nu}(k).\end{eqnarray}
\end{center}
\end{figure}
\newline
Here the blobs mean the sum of the one-particle irreducible diagrams. In the standard way one shows that $\Pi^{(A)}_{\mu\nu}$ and $\Pi^{(G)}_{\mu\nu}$ are purely transverse and independent of the ``gauge parameters'' $\eta_{0}$ and $\xi_{0}$. Thus we write
\begin{eqnarray}
\Pi^{(A,G)}_{\mu\nu}(k)=(-g_{\mu\nu}k^2+k_{\mu}k_{\nu})\Pi^{(A,G)}(k^2).
\end{eqnarray}
The unrenormalised photon propagator $\Delta^{'(A)}_{F\mu\nu}$ reads
\begin{eqnarray}
\label{eqn32000}
i\Delta^{'(A)}_{F}(k)_{\mu\nu}&=&\frac{-i}{k^2[1+e_{0}^2\Pi^{(A)}(k^2)]-\lambda_{0}^2+i\epsilon}\nonumber\\
&&\qquad[g_{\mu\nu}+(\xi_{0}[1+e_{0}^2\Pi^{(A)}(k^2)]-1)\frac{k_{\mu}k_{\nu}}{k^2-\xi_{0}\lambda_{0}^2+i\epsilon}].
\end{eqnarray}
The renormalised photon mass $\lambda$ is obtained from
\begin{eqnarray}
\lambda^2[1+e_{0}^2\Pi^{(A)}(\lambda^2)]-\lambda_{0}^2=0.
\end{eqnarray}
The wave function renormalisation constant $Z_{3}^{(A)}$ of the photon is given by
\begin{eqnarray}
\label{eqn30000}
Z_{3}^{(A)}&=&[1+e_{0}^2\Pi^{(A)}(\lambda^2)]^{-1}.
\end{eqnarray}
It follows that the renormalised mass $\lambda$, the renormalised charge $e$ and the renormalised gauge parameter $\xi$ are given by
\begin{eqnarray}
\lambda^2&=&Z_{3}^{(A)}\lambda_{0}^2,\\
e^2&=&Z_{3}^{(A)}e_{0}^2,\\
\label{eqn10000}
\xi&=&(Z_{3}^{(A)})^{-1}\xi_{0}.
\end{eqnarray}
With that we get for the renormalised field $A^{\mu}_{R}$ and propagator $\Delta^{(A)}_{F,R\mu\nu}$
\begin{eqnarray}
A^{\mu}_{R}(x)&=&(Z_{3}^{(A)})^{-\frac{1}{2}}A^{\mu}(x),\\
i\Delta^{(A)}_{F,R}(k)_{\mu\nu}&=&\frac{-i}{k^2[1+e^2\Pi_{c}^{(A)}(k^2)]-\lambda^2+i\epsilon}\nonumber\\
&&\qquad[g_{\mu\nu}+(\xi[1+e^2\Pi^{(A)}_{c}(k^2)]-1)\frac{k_{\mu}k_{\nu}}{k^2-\xi\lambda^2+i\epsilon}]\\
\Pi_{c}^{(A)}(k^2)&=&\Pi^{(A)}(k^2)-\Pi^{(A)}(\lambda^2)\label{eqn33000}.
\end{eqnarray}
Note that with the conventional choice of $Z_{3}^{A}$ in (\ref{eqn30000}) the residue of the pole at $k^2=\lambda^2$ in the transverse term of the renormalised propagator is finite but not normalised to 1.

\medskip

For the gluon relations completely analogous to (\ref{eqn32000}) - (\ref{eqn33000}) hold with the replacements $A\to G$, $e_{0}\to g_{0}$, $\lambda_{0}\to\mu_{0}$ and $\xi_{0}\to\eta_{0}$.

\medskip

The renormalisation of the electron and quark fields is standard. Let $-i\Sigma^{(\psi)}(p)$ be the unrenormalised electron self energy, then the unrenormalised electron propagator $S_{F}^{'(\psi)}$ is
\begin{eqnarray}
\label{eqn34000}
S_{F}^{'(\psi)}(p)^{-1}=\pslash-m_{0}-\Sigma^{(\psi)}(p)+i\epsilon.
\end{eqnarray}
The renormalised electron mass $m$ and the wave function renormalisation constant $Z_{2}^{(\psi)}$ are given by
\begin{eqnarray}
0&=&m-m_{0}-\Sigma(m),\\
Z_{2}^{(\psi)}&=&[1-\frac{\partial}{\partial\pslash}\ \Sigma^{(\psi)}(\pslash\  )\raisebox{-1ex}{\em $\mid_{p\!\!\!\!\not\,\, \ =m}$}]^{-1}.
\end{eqnarray}
For the renormalised electron field $\psi_{R}$ and propagator $S_{F,R}^{(\psi)}$ we have
\begin{eqnarray}
\psi_{R}(x)&=&(Z_{2}^{(\psi)})^{-\frac{1}{2}}\psi(x),\\
S_{F,R}^{(\psi)}(p)^{-1}&=&\pslash-m-\Sigma^{(\psi)}_{R}(p)+i\epsilon,\\
\Sigma^{(\psi)}_{R}(p)&=&Z_{2}^{(\psi)}[\Sigma^{(\psi)}(p)-\Sigma^{(\psi)}(m)]-(Z_{2}^{(\psi)}-1)(\pslash-m)\label{eqn35000}.
\end{eqnarray}

\medskip

Relations analogous to (\ref{eqn34000}) - (\ref{eqn35000}) hold for quarks in our theory with the replacements $\psi\to q$, $m_{0}\to M_{0}$. Note that also the complete quark propagator is flavour diagonal due to the $C_{A}$ and $C_{G}$ symmetries (\ref{eqn36000}),(\ref{eqn37000}).

\medskip
 
This concludes our brief discussion of the renormalisation of the model.

\section{Free Fields and Time-shift Relations}

\setcounter{equation}{0}

\subsection{Massive Vector Field}

A free vector field $G_{\mu}$ of mass $\mu$ can be described by a Lagrangian density
\begin{eqnarray}
{\cal L}=-\frac{1}{4}G_{\mu\nu}G^{\mu\nu}-\frac{1}{2\eta}(\partial_{\mu}G^{\mu})^2+\frac{1}{2}\mu^2G_{\mu}G^{\mu}
\end{eqnarray}
with the field strength tensor $G_{\mu\nu}=\partial_{\mu}G_{\nu}-\partial_{\nu}G_{\mu}$. With the canonical momenta $\Pi^{\epsilon}$
\begin{eqnarray}\label{eqn9000}
\Pi^{\epsilon}=\frac{\partial{\cal L}}{\partial(\partial_{0}G_{\epsilon})}=G^{\epsilon 0}-g^{\epsilon 0}\frac{1}{\eta}(\partial_{\mu}G^{\mu}),
\end{eqnarray}
the Hamiltonian density is given as
\begin{eqnarray}
{\cal H}=-\frac{1}{2}\eta\Pi_{0}\Pi^{0}-\frac{1}{2}\Pi_{i}\Pi^{i}-\partial_{i}G^{i}\Pi^0+\Pi^{i}\partial_{i}G^0+\frac{1}{4}G_{ij}G^{ij}-\frac{1}{2}\mu^2G_{\mu}G^{\mu}.
\end{eqnarray}
Using the commutator function $\Delta$
\begin{eqnarray}
\Delta(x ;\mu^2)=i\int\! d^{R}k (e^{-ikx}-e^{ikx})
\end{eqnarray}
with
\begin{eqnarray}
d^{R}k =\frac{d^4\!k}{(2\pi)^3}\theta(k^0)\delta^{(1)}(k^2-\mu^2),
\end{eqnarray}
$G_{\mu}$ at the space-time $x$ can be obtained from $G_{\mu}$ at time $y^0$ by means of
\begin{eqnarray}
\label{eqn50}
G_{\mu}(x) 
&=& \int_{y^0=const.} d^3y\ [\ \dot{\Delta}(x-y;\mu^{2})G_{\mu}(y)-\Delta(x-y;\mu^{2})\Pi_{\mu}(y)+\nonumber\\
& &\qquad\qquad\qquad+\frac{1}{\mu^{2}}\partial_{\mu}\partial_{\rho}[\Delta(x-y;\eta\mu^2)-\Delta(x-y;\mu^{2})]\Pi^{\rho}(y)\nonumber\\
& &\qquad\qquad\qquad+\partial_{\mu}\Delta(x-y;\eta\mu^{2})G^{0}(y)-g_{0\mu}\partial_{\rho}\Delta(x-y;\mu^{2})G^{\rho}(y)\ ],\nonumber\\
\end{eqnarray}
where the time-derivative of $G_{\mu}$ has been expressed in terms of $G_{\mu}$ and $\Pi_{\mu}$ with the help of the free equations of motion. The derivatives are taken with respect to the first argument. In the case of $\eta=1$ (\ref{eqn50}) reduces to
\begin{eqnarray}
\label{eqn80}
G_{\mu}(x)&=&\int_{y^0=const.}d^3y\ \Delta(x-y;\mu^2)\partial_{y^{0}}\!\!\!\!\!\!\!\!\raisebox{1.5ex}{\em$\leftrightarrow$}\ G_{\mu}(y).
\end{eqnarray}
Furthermore we need for the perturbative calculation the propagator
\begin{eqnarray}
\langle 0|\ \mbox{T} G_{\mu}(x)G_{\rho}(y)\ | 0 \rangle=\int\frac{d^4\!k}{(2\pi)^4}\frac{-ig_{\mu\rho}}{k^2-\mu^2+i\epsilon}e^{-ik(x-y)}.
\end{eqnarray}

\subsection{Dirac Field}

Starting from the Lagrangian density for a free Dirac field $\psi$ of mass $m$
\begin{eqnarray}
{\cal L}=\ \bar{\psi}\left(\frac{i}{2}\!\!\!\!\stackrel{\ \ \leftrightarrow}{\partialslash}-m\right)\psi,
\end{eqnarray}
the corresponding Hamiltonian density is given as
\begin{eqnarray}
{\cal H}=\ -\bar{\psi}\left(\frac{i}{2}\gamma^{i}\!\!\!\!\stackrel{\ \ \leftrightarrow}{\ \partial_{i}}-m\right)\psi.
\end{eqnarray}
The mapping of $\psi$ from $y$ to another space-time point $x$ is accomplished by
\begin{eqnarray}
\label{eqn60}
\psi(x)&=&-i\int_{y^0=const.}d^3y S(x-y; m)\gamma^0\psi(y)\\
\bar{\psi}(x)&=&-i\int_{y^0=const.}d^3y\ \bar{\psi}(y)\gamma^0 S(y-x; m)
\end{eqnarray}
with
\begin{eqnarray}
S(x; m) =(i\partialslash_{x}+m)\Delta(x ;m^2).
\end{eqnarray}
Furthermore we need the perturbative propagator
\begin{eqnarray}
\label{eqn100}
\langle 0 |\ \mbox{T}\psi(x)\bar{\psi}(y)\ | 0 \rangle&=&\int\! d^{R}p\ [\theta(x^0-y^0)e^{-ip(x-y)}(\pslash+m)\nonumber\\
& &\qquad\qquad+\theta(y^0-x^0)e^{+ip(x-y)}(-\pslash+m)].
\end{eqnarray}

\section{Calculation of ${\cal M}$}

\setcounter{equation}{0}

The calculation of ${\cal M}$ is demonstrated best in calculating the part 
\begin{eqnarray}
I&=&\langle\!\langle 0 |\ \mbox{T}\int\! d^4\!x_{3}[\theta(-x_{3}^0)A_{\mu}(x_{3})e^{i(p_{1}+p_{2})x_{3}}-\theta(x_{3}^0)A_{\mu}(\tilde{x_{3}})e^{i(p_{1}+p_{2})\tilde{x_{3}}}]\nonumber\\
&&\qquad\qquad\int\! d^4\!x_{1}[\theta(-x_{1}^0)\bar{q}(x_{1})\Aslash(x_{1})q(x_{1})-\theta(x_{1}^0)\bar{q}(\tilde{x}_{1})\Aslash(\tilde{x}_{1})q(\tilde{x}_{1})]\ ...\ | 0 \rangle\!\rangle\nonumber\\
\end{eqnarray}
of the matrix element in (\ref{eqn500000}). This expression is in principle governed by the four integrals $I_{1}$, $I_{2}$, $I_{3}$ and $I_{4}$, which will be calculated successively in the following. We will start with  $I_{1}$
\begin{eqnarray}
I_{1}&=&\langle\!\langle 0 |\ \mbox{T}\int\! d^4\!x_{3}\theta(-x_{3}^0)A_{\mu}(x_{3})e^{i(p_{1}+p_{2})x_{3}}\theta(-x_{1}^0)A_{\rho}(x_{1})\ | 0 \rangle\!\rangle\nonumber\\
&=&\theta(-x_{1}^0)\int\! d^4\!x_{3}\theta(-x_{3}^0)e^{i(p_{1}+p_{2})x_{3}}\langle\!\langle 0 |\ \mbox{T}A_{\mu}(x_{3})A_{\rho}(x_{1})\ | 0 \rangle\!\rangle\nonumber\\
&=&-ig_{\mu\rho}\theta(-x_{1}^0)\int\! d^4\!x_{3}\theta(-x_{3}^0)e^{i(p_{1}+p_{2})x_{3}}\int\frac{d^4\!k}{(2\pi)^4}\frac{1}{k^2-\lambda^2+i\epsilon}e^{-ik(x_{3}-x_{1})}\nonumber\\
&=&-ig_{\mu\rho}\theta(-x_{1}^0)\int d^3\!k\ \delta^{(3)}(\vec{p_{1}}+\vec{p_{2}}-\vec{k})\int\! dx_{3}^0\theta(-x_{3}^0)e^{i(p_{1}^0+p_{2}^0)x_{3}^0}\nonumber\\
&&\qquad\int\frac{dk^0}{(2\pi)}\frac{1}{k^2-\lambda^2+i\epsilon}e^{-ik^0(x_{3}^0-x_{1}^0)}.
\end{eqnarray}
The integration of $k^0$ is best done using the residue theorem. Unfortunately, because of the occurring $\theta$-functions in $x_{3}^0$ and $x_{1}^0$ the integral is neither in the upper nor in the lower complex plane convergent. This can be circumvented via 
\begin{eqnarray}
\label{eqn2000}
\int\! d^4\!x_{3}\theta(-x_{3}^0)f(x_{3})=\int\! d^4\!x_{3}f(x_{3})-\int\! d^4\!x_{3}\theta(x_{3}^0)f(x_{3})
\end{eqnarray}
so that we get two contributions, one in which the $x_{3}^0$-integration stretches over the whole real axis, we get a $\delta$-function in $k^0$ and its integration becomes trivial and a second contribution where we can apply the residue theorem without any problems  
\begin{eqnarray}
I_{1}&=&-ig_{\mu\rho}\theta(-x_{1}^0)[\frac{1}{s-\lambda^2+i\epsilon}e^{i(p_{1}+p_{2})x_{1}}\nonumber\\
&&\qquad\qquad\qquad-\int d^{R}k(2\pi)^3\delta^{(3)}(\vec{p_{1}}+\vec{p_{2}}-\vec{k})e^{ikx_{1}}\frac{1}{p_{1}^0+p_{2}^0-k^{0}+i\epsilon}].\nonumber\\&&
\end{eqnarray}
With the help of
\begin{eqnarray}
\label{eqn5000}
&&\int d^4\!B\delta^{(1)}(B^0-x_{3}^0)\Delta(\tilde{x_{3}}-B)\partial_{B^{0}}\!\!\!\!\!\!\!\!\raisebox{1.5ex}{\em$\leftrightarrow$}\ e^{-ikB} =\nonumber\\
&&\ \ = e^{-ikx_{3}}\frac{1}{2k^{0'}}\left( e^{2ik^{0'}x_{3}^0}(k^0+k^{0'})+(-k^0+k^{0'}) e^{-2ik^{0'}x_{3}^0}\right)\ \ \mbox{with}\ k^{0'}=\sqrt{\vec{k}^2+\lambda^2}\nonumber\\&&
\end{eqnarray}
the integration of $I_{2}$ is straightforward. We can apply the residue theorem directly because the $\theta$-functions ensure the convergence of the integral over $k^0$
\begin{eqnarray}
I_{2}&=&\langle\!\langle 0 |\ \mbox{T}\int\! d^4\!x_{3}\theta(x_{3}^0)A_{\mu}(\tilde{x_{3}})e^{i(p_{1}+p_{2})\tilde{x_{3}}}\theta(-x_{1}^0)A_{\rho}(x_{1})\ | 0 \rangle\!\rangle\nonumber\\
&=&\theta(-x_{1}^0)\int\! d^4\!x_{3}\theta(x_{3}^0)e^{i(p_{1}+p_{2})\tilde{x_{3}}}\langle\!\langle 0 |\ \mbox{T}A_{\mu}(\tilde{x_{3}})A_{\rho}(x_{1})\ | 0 \rangle\!\rangle\nonumber\\
&=&\theta(-x_{1}^0)\int\! d^4\!x_{3}\theta(x_{3}^0)e^{i(p_{1}+p_{2})\tilde{x_{3}}}\int d^4\!B\delta^{(1)}(B^0-x_{3}^0)\nonumber\\
&&\qquad\qquad\Delta(\tilde{x_{3}}-B)\partial_{B^{0}}\!\!\!\!\!\!\!\!\raisebox{1.5ex}{\em$\leftrightarrow$}\ \langle\!\langle 0 |\ \mbox{T}A_{\mu}(B)A_{\rho}(x_{1})\ | 0 \rangle\!\rangle\nonumber\\
&=&ig_{\mu\rho}\theta(-x_{1}^0)\int d^{R}k(2\pi)^3\delta^{(3)}(\vec{p_{1}}+\vec{p_{2}}-\vec{k})e^{ikx_{1}}\frac{1}{p_{1}^0+p_{2}^0-k^{0}-i\epsilon}.
\end{eqnarray}
To calculate $I$ we need only the difference between both integrals
\begin{eqnarray}
I_{1}-I_{2}&=&-ig_{\mu\rho}\theta(-x_{1}^0)e^{i(p_{1}+p_{2})x_{1}}\nonumber\\
&&\qquad\qquad[\frac{1}{s-\lambda^2+i\epsilon}+i\int d^{R}k(2\pi)^4\delta^{(4)}(p_{1}+p_{2}-k)]\nonumber\\
&=&-ig_{\mu\rho}\theta(-x_{1}^0)\frac{1}{s-\lambda^2-i\epsilon}e^{i(p_{1}+p_{2})x_{1}}.
\end{eqnarray}
$I_{3}$ can be treated in the same way as $I_{2}$
\begin{eqnarray}
I_{3}&=&\langle\!\langle 0 |\ \mbox{T}\int\! d^4\!x_{3}\theta(-x_{3}^0)A_{\mu}(x_{3})e^{i(p_{1}+p_{2})x_{3}}\theta(x_{1}^0)A_{\rho}(\tilde{x}_{1})\ | 0 \rangle\!\rangle\nonumber\\
&=&ig_{\mu\rho}\theta(x_{1}^0)\int d^{R}k(2\pi)^3\delta^{(3)}(\vec{p_{1}}+\vec{p_{2}}-\vec{k})e^{ikx_{1}}\frac{1}{p_{1}^0+p_{2}^0+k^{0'}-i\epsilon}.
\end{eqnarray} 
The most complicated integral is $I_{4}$
\begin{eqnarray}
I_{4}&=&\langle\!\langle 0 |\ \mbox{T}\int\! d^4\!x_{3}\theta(x_{3}^0)A_{\mu}(\tilde{x_{3}})e^{i(p_{1}+p_{2})\tilde{x_{3}}}\theta(x_{1}^0)A_{\rho}(\tilde{x}_{1})\ | 0 \rangle\!\rangle\nonumber\\
&=&\theta(x_{1}^0)\int\! d^4\!x_{3}\theta(x_{3}^0)e^{i(p_{1}+p_{2})\tilde{x_{3}}}\langle\!\langle 0 |\ \mbox{T}A_{\mu}(\tilde{x_{3}})A_{\rho}(\tilde{x}_{1})\ | 0 \rangle\!\rangle\nonumber\\
&=&\theta(x_{1}^0)\int\! d^4\!x_{3}\theta(x_{3}^0)e^{i(p_{1}+p_{2})\tilde{x_{3}}}\int d^4\!B\delta^{(1)}(B^0-x_{3}^0)\Delta(\tilde{x_{3}}-B)\nonumber\\
&&\qquad\int d^4\!C\delta^{(1)}(C^0-x_{1}^0)\Delta(\tilde{x_{1}}-C)\langle\!\langle 0 |\ \mbox{T}\partial_{B^{0}}\!\!\!\!\!\!\!\!\raisebox{1.5ex}{\em$\leftrightarrow$}\ A_{\mu}(B)\partial_{C^{0}}\!\!\!\!\!\!\!\!\raisebox{1.5ex}{\em$\leftrightarrow$}\ A_{\rho}(C)\ | 0 \rangle\!\rangle.
\end{eqnarray}
From this integral we get two contributions because of an additional contact term which arises due to the two time derivatives in the 2-point function
\begin{eqnarray}
&&\langle\!\langle 0 |\ \mbox{T}\partial_{B^{0}}A_{\mu}(B)\partial_{C^{0}}A_{\rho}(C)\ | 0 \rangle\!\rangle =\nonumber\\
&&\qquad\qquad = \partial_{B^{0}}\partial_{C^{0}}\langle\!\langle 0 |\ \mbox{T}A_{\mu}(B)A_{\rho}(C)\ | 0 \rangle\!\rangle +ig_{\mu\rho}\delta^{(4)}(B-C).
\end{eqnarray}
The contact term gives rise to the contribution
\begin{eqnarray}
I_{41}&=&-ig_{\mu\rho}\theta(x_{1}^0)e^{i(p_{1}+p_{2})\tilde{x}_{1}}\nonumber\\
&&\qquad\qquad\qquad\int d^{R}k\frac{1}{2k^0}(2\pi)^3\delta^{(3)}(\vec{p_{1}}+\vec{p_{2}}-\vec{k})(e^{4ik^0x_{1}^0}+e^{-4ik^0x_{1}^0}-2)\nonumber\\ 
\end{eqnarray}
and the rest, again with the use of (\ref{eqn2000}) and (\ref{eqn5000}), results into
\begin{eqnarray}
I_{42}&=&ig_{\mu\rho}\theta(x_{1}^0)[\int d^{R}k(2\pi)^3\delta^{(3)}(\vec{p_{1}}+\vec{p_{2}}-\vec{k})e^{ikx_{1}}\frac{1}{p_{1}^0+p_{2}^0+k^{0'}+i\epsilon}\nonumber\\
&&\qquad\qquad-e^{i(p_{1}+p_{2})\tilde{x}_{1}}(\int d^{R}k\frac{1}{2k^0}(2\pi)^3\delta^{(3)}(\vec{p_{1}}+\vec{p_{2}}-\vec{k})(2-e^{4ik^0x_{1}^0}-e^{-4ik^0x_{1}^0})\nonumber\\
&&\qquad\qquad\qquad\qquad\qquad\qquad-\frac{1}{s-\lambda^2-i\epsilon})],
\end{eqnarray}
so that we get
\begin{eqnarray}
I_{4}&=&ig_{\mu\rho}\theta(x_{1}^0)[\int d^{R}k(2\pi)^3\delta^{(3)}(\vec{p_{1}}+\vec{p_{2}}-\vec{k})e^{ikx_{1}}\frac{1}{p_{1}^0+p_{2}^0+k^{0'}+i\epsilon}\nonumber\\
&&\qquad\qquad+e^{i(p_{1}+p_{2})\tilde{x}_{1}}\frac{1}{s-\lambda^2-i\epsilon}].
\end{eqnarray}
Again we only need the difference between both integrals
\begin{eqnarray}
I_{3}-I_{4}&=&ig_{\mu\rho}\theta(x_{1}^0)e^{i(p_{1}+p_{2})\tilde{x}_{1}}\nonumber\\
&&\qquad\qquad[-\frac{1}{s-\lambda^2-i\epsilon}+i\int d^{R}k(2\pi)^4\delta^{(4)}(p_{1}+p_{2}+k)]\nonumber\\
&=&-ig_{\mu\rho}\theta(x_{1}^0)\frac{1}{s-\lambda^2-i\epsilon}e^{i(p_{1}+p_{2})\tilde{x}_{1}},
\end{eqnarray}
so that we finally get
\begin{eqnarray}
&&I=-i\frac{1}{s-\lambda^2-i\epsilon}\nonumber\\
&&\ \ \langle\!\langle 0 |\mbox{T}\int\! d^4\!x_{1}[\theta(-x_{1}^0)\bar{q}(x_{1})\gamma_{\mu}q(x_{1})e^{i(p_{1}+p_{2})x_{1}}-\theta (x_{1}^0)\bar{q}(\tilde{x}_{1})\gamma_{\mu}q(\tilde{x}_{1})e^{i(p_{1}+p_{2})\tilde{x}_{1}}] ... | 0 \rangle\!\rangle.\nonumber\\
\end{eqnarray}

\end{fmffile}

\end{document}